\documentclass[fleqn,usenatbib]{mnras}

\usepackage{newtxtext,newtxmath}
\usepackage{amsmath,amssymb}

\usepackage[T1]{fontenc}
\usepackage{ae,aecompl}

\usepackage{graphicx}	% Including figure files
\usepackage{amsmath}	% Advanced maths commands
\usepackage{amssymb}	% Extra maths symbols

\usepackage{xcolor}
\usepackage{footnote}

\newcommand{\refeq}[1]{Eq.~(\ref{eq:#1})}          
          
\newcommand{\reffig}[1]{Fig.~\ref{fig:#1}}          
\newcommand{\refsec}[1]{Sec.~\ref{sec:#1}}
\newcommand{\refapp}[1]{App.~\ref{app:#1}}
\newcommand{\reftab}[1]{Tab.~\ref{tab:#1}}

\def\rmd{\mathrm{d}}
\def\bfx{\boldsymbol{x}}
\def\bfp{\boldsymbol{p}}

\begin{document}
\label{firstpage}
\pagerange{\pageref{firstpage}--\pageref{lastpage}}

% Title of the paper, and the short title which is used in the headers.
% Keep the title short and informative.
\title[Asymmetry of Arc in SDSS J1226+2152]{Asymmetric Surface Brightness Structure of Caustic Crossing Arc in SDSS J1226+2152: A Case for Dark Matter Substructure}
 
% The list of authors, and the short list which is used in the headers.
% If you need two or more lines of authors, add an extra line using \newauthor
\author[L. Dai et al.]{
Liang Dai,$^{1}$\thanks{E-mail: ldai@ias.edu}
Alexander A. Kaurov,$^{1}$
Keren Sharon,$^{2}$
Michael Florian,$^{3}$
\newauthor{Jordi Miralda-Escud\'{e},$^{4,5}$
Tejaswi Venumadhav,$^{1}$
Brenda Frye,$^{6}$}
\newauthor{Jane R. Rigby,$^{3}$
and Matthew Bayliss$^{7,8}$}
\\
% List of institutions
$^{1}$Institute for Advanced Study, 1 Einstein Drive, Princeton, NJ 08540, USA\\
$^{2}$Department of Astronomy, University of Michigan, 1085 S. University Ave, Ann Arbor, MI 48109, USA\\
$^{3}$Observational Cosmology Lab, NASA Goddard Space Flight Center, 8800 Greenbelt Rd., Greenbelt, MD 20771, USA\\
$^{4}$Instituci\'o Catalana de Recerca i Estudis Avan\c cats, Barcelona, Catalonia\\
$^{5}$Institut de Ci\`encies del Cosmos, Universitat de Barcelona (IEEC-UB), Barcelona, Catalonia\\
$^{6}$Department of Astronomy/Steward Observatory, University of Arizona, 933 N. Cherry Ave, Tucson, AZ  85721, USA\\
$^{7}$Department of Physics, University of Cincinnati, Cincinnati, OH 45221, USA \\
$^{8}$MIT Kavli Institute for Astrophysics and Space Research, 77 Massachusetts Ave., Cambridge, MA 02139, USA
}

% These dates will be filled out by the publisher
\date{Accepted XXX. Received YYY; in original form ZZZ}

% Enter the current year, for the copyright statements etc.
%\pubyear{2019}

\maketitle

%%%%%%%%%%%%%%%%%%%%%%%%%%%%%%%%%%%%%%%%%%%%%%%%%%%%%%%%%%%%%%%%%%%%%%%%%%%%%%%%%%%
% Abstract of the paper
\begin{abstract}

We study the highly magnified arc SGAS J122651.3+215220 caused by a star-forming galaxy at $z_s=2.93$ crossing the lensing caustic cast by the galaxy cluster SDSS J1226+2152 ($z_l=0.43$), using Hubble Space Telescope observations. We report in the arc several asymmetric surface brightness features whose angular separations are a fraction of an arcsecond from the lensing critical curve and appear to be highly but unequally magnified image pairs of underlying compact sources, with one brightest pair having clear asymmetry consistently across four filters. One explanation of unequal magnification is microlensing by intracluster stars, which induces independent flux variations in the images of individual or groups of source stars in the lensed galaxy. For a second possibility, intracluster dark matter subhalos invisible to telescopes effectively perturb lensing magnifications near the critical curve and give rise to persistently unequal image pairs. Our modeling suggests, at least for the most prominent identified image pair, that the microlensing hypothesis is in tension with the absence of notable asymmetry variation over a six-year baseline, while subhalos of $\sim 10^6$--$10^8\,M_\odot$ anticipated from structure formation with Cold Dark Matter typically produce stationary and sizable asymmetries. We judge that observations at additional times and more precise lens models are necessary to stringently constrain temporal variability and robustly distinguish between the two explanations. The arc under this study is a scheduled target of a Director's Discretionary Early Release Science program of the James Webb Space Telescope, which will provide deep images and a high-resolution view with integral field spectroscopy.

\end{abstract}
%%%%%%%%%%%%%%%%%%%%%%%%%%%%%%%%%%%%%%%%%%%%%%%%%%%%%%%%%%%%%%%%%%%%%%%%%%%%%%%%%%%

% Select between one and six entries from the list of approved keywords.
% Don't make up new ones.

%%%%%%%%%%%%%%%%%%%%%%%%%%%%%%%%%%%%%%%%%%%%%%%%%%

%%%%%%%%%%%%%%%%% BODY OF PAPER %%%%%%%%%%%%%%%%%%

%%%%%%%%%%%%%%%%%%%%%%%%%%%%%%%%%%%%%%%%%%%%%%%%%%%%%%%%%%%%%%%%%%%%
\section{Introduction}
%%%%%%%%%%%%%%%%%%%%%%%%%%%%%%%%%%%%%%%%%%%%%%%%%%%%%%%%%%%%%%%%%%%%

Massive clusters of galaxies are powerful cosmic gravitational lenses that often make outstanding magnified images of background galaxies, usually referred to as ``arcs''. The most dramatic phenomenon takes place when a background galaxy happens to straddle the caustic cast by the foreground gravitational lens, offering an opportunity to closely study an elongated image of the galaxy with a zoomed-in view near the caustic. Smoothly distributed mass in the foreground lens universally produces symmetric pairs of merging images near a fold lensing critical curve. However, intracluster microlensing of individual superluminous stars in the source galaxy~\citep{1991ApJ...379...94M, 2017ApJ...850...49V, 2018ApJ...857...25D, Oguri:2017ock} or sub-galactic dark matter (DM) substructure predicted to exist in cluster halos~\citep{press1974formation, white1978core,  1985ApJ...292..371D}, can cause temporary or persistent departure from the symmetric appearance of the arc, respectively.

By analysing this image asymmetry one can study distant extragalactic objects that are otherwise prohibitively difficult to probe. First, microlensing allows one to scrutinize individual bright stars at $z \gtrsim 1$~\citep{2018NatAs...2..334K, Chen:2019ncy, 2019ApJ...880...58K}, or reveal stellar transients from the distant Universe~\citep{2018NatAs...2..324R, Diego:2018fzr, windhorst2018observability}. The upcoming 192-orbit HST program (GO-15936, PI: Kelly) may increase the number of detections of extremely bright extragalactic stars~\citep{kelly2019flashlights}. This may be used to probe the intracluster population of stellar microlenses inside the cluster halo, or even a multitude of (sub-)planetary mass self-gravitating DM minihalos if they form in large abundance during the early Universe~\citep{Dai:2019lud, Arvanitaki:2019rax, Blinov:2019jqc}. Second, perturbations in the shape of the critical curve on scales $\sim 0.1\arcsec$ will inform us about the abundance of  $10^6-10^8\,M_\odot$ DM subhalos within the cluster-sized $10^{13}-10^{15}\,M_\odot$ lens halo~\citep{2018ApJ...867...24D}, thereby probing a dynamic range not easily accessible with other methods.

Previously, multi-epoch Hubble Space Telescope (HST) observations in the field of the galaxy cluster MACS J0416.1-2403 allowed \cite{Chen:2019ncy} and \cite{2019ApJ...880...58K} to carry out independent studies of time-varying image asymmetries in a $z=0.9397$ arc crossing the cluster critical curve.  Both groups suggested that the most significant asymmetric feature is at least partially caused by microlensing induced random brightening of macro images of an underlying supergiant star. Interestingly,
the boost in brightness was detected in 2014 for the former and in 2012 for the latter. Moreover, \cite{2019ApJ...880...58K} mentioned that the arc may have several other less significant asymmetric features. 

In this paper, we examine the giant arc SGAS J122651.3+215220 (hereafter S1226) with a spectroscopic redshift $z_s=2.93$ in the field of the $z = 0.43$ galaxy cluster SDSS J1226+2152. This arc, resulting from two images of a background galaxy merging along a critical line, was first discovered from the SDSS Giant Arcs Survey (SGAS) as a Lyman Break Galaxy during intense star formation~\citep{koester2010two}. Concerning the study of image asymmetries, S1226 is particularly advantageous because it exhibits multiple bright compact star-forming clumps, whose image pairs across the critical curve have small angular separations $\lesssim 0.3\arcsec$ and hence are highly magnified. We examine HST imaging data obtained in 2011 in two optical filters F606W and F814W, and data obtained in 2017 in two infrared (IR) filters F110W and F160W.
We identify several asymmetric image pairs following a method similar to the one presented in \cite{2019ApJ...880...58K} for an arc in the field of MACS J0416.1-2403, and then carry out an analysis based on template matching to robustly measure the absolute fluxes and flux asymmetry of image pairs.

We identify in S1226 at least three magnified image pairs of significant surface brightness asymmetries in the F606W and F814W filters, among which the brightest one also shows significant and consistent asymmetry in the lower resolution F110W and F160W filters. This is unexpected for a smooth mass distribution in the lens. Located within a fraction of an arcsecond from the critical curve, these asymmetric features must be highly magnified images of underlying sources in the star-forming galaxy smaller than a few parsecs, and the high intrinsic luminosities imply that they are probably groups of many bright stars rather than individual super-bright stars. The brightest pair has a high degree of asymmetry consistent in all four HST filters, with tentative evidence for persistence over a six-year observing baseline. We find that the absence of significant variability in asymmetric features implies tension with stochastic intracluster microlensing as the explanation for the asymmetry. On the other hand, we suggest that lensing by sub-galactic DM subhalos of masses $\sim 10^6$--$10^8\,M_\odot$ as predicted by the standard theory of hierarchical structure assembly with Cold Dark Matter (CDM)~\citep{moore1999dark, 2008MNRAS.391.1685S, 2016ApJ...824..144F} can explain persistent asymmetries at the observed level. This derives from the idea of using image flux ratios as diagnostic of invisible substructure~\citep{1998MNRAS.295..587M, 2001ApJ...563....9M}, but extends from galactic host halos~\citep{2019MNRAS.487.5721G} to more massive ones $M_{\rm host} \gtrsim 10^{14}\,M_\odot$ and into the regime of large magnification factors $\gtrsim 100$, by examining just one caustic straddling system. Lack of existing multi-epoch imaging in any single HST filter, however, prevents us from decisively ruling out microlensing variability in favor of substructure lensing. Promisingly, S1226 has been selected as a target of the approved Director's Discretionary Early Release Science (DD-ERS) program with the James Webb Space Telescope (JWST) (\# 1335, PI: Rigby)~\citep{rigby2017templates}. Forthcoming data will make this highly magnified arc a particularly interesting system.

This paper is organized as follows: in \refsec{data} we summarize existing HST and the upcoming JWST data. Then in \refsec{asym}, we show the under-resolved star-forming clumps exhibiting significant flux asymmetry between their highly magnified image pairs. In \refsec{macrolens}, we use the macro lens model describing the galaxy cluster SDSS J1226+2152 to infer coarse-grained properties at the location of the arc. In \refsec{icstars}, we estimate the abundance of foreground intracluster stars in the arc projected vicinity, which is useful information for quantitatively understanding microlensing. We then consider two possible physical explanations for the observed image asymmetry: time-varying intracluster microlensing in \refsec{microlensing}, and persistent DM substructure lensing in \refsec{subhalolensing}. We conclude that intracluster microlensing is not conclusively ruled out but is unfavorable in light of the roughly consistent asymmetries across a six-year observational baseline, while DM subhalo lensing is theoretically expected and naturally explains non-variability. Finally, we give concluding remarks in \refsec{disc}. Furthermore, we include \refapp{nonrig} presenting non-rigid registration as an alternative method to corroborate the existence of flux asymmetries, and \refapp{analysisdetails} providing data analysis details. In \refapp{magasym}, we analytically estimate magnification asymmetry induced by perturber lenses in the proximity of the critical curve, supporting the argument that nearby minor cluster member galaxies are not the cause of the observed asymmetries.

%%%%%%%%%%%%%%%%%%%%%%%%%%%%%%%%%%%%%%%%%%%%%%%%%%%%%%%%%%%%%%%%%%%%
\begin{figure*}
	% To include a figure from a file named example.*
	% Allowable file formats are eps or ps if compiling using latex
	% or pdf, png, jpg if compiling using pdflatex
	\includegraphics[scale=0.8]{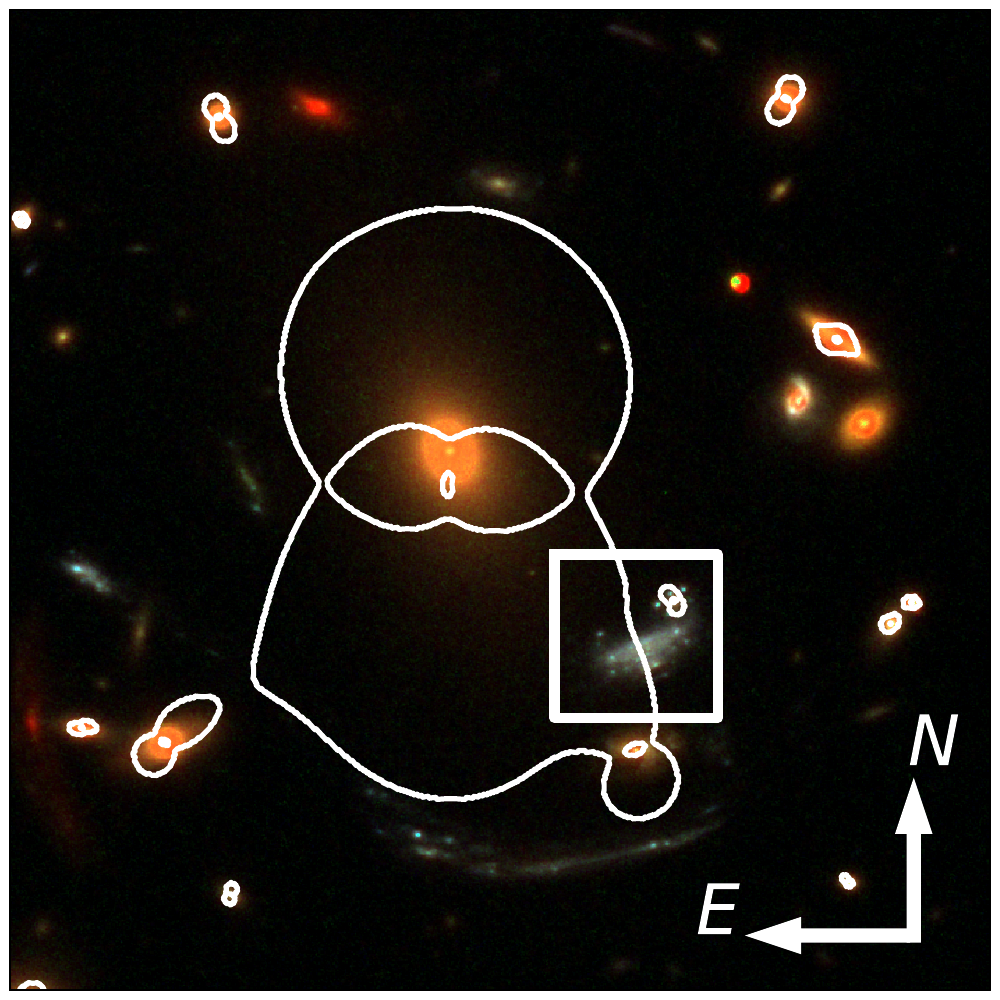}
	\includegraphics[scale=0.8]{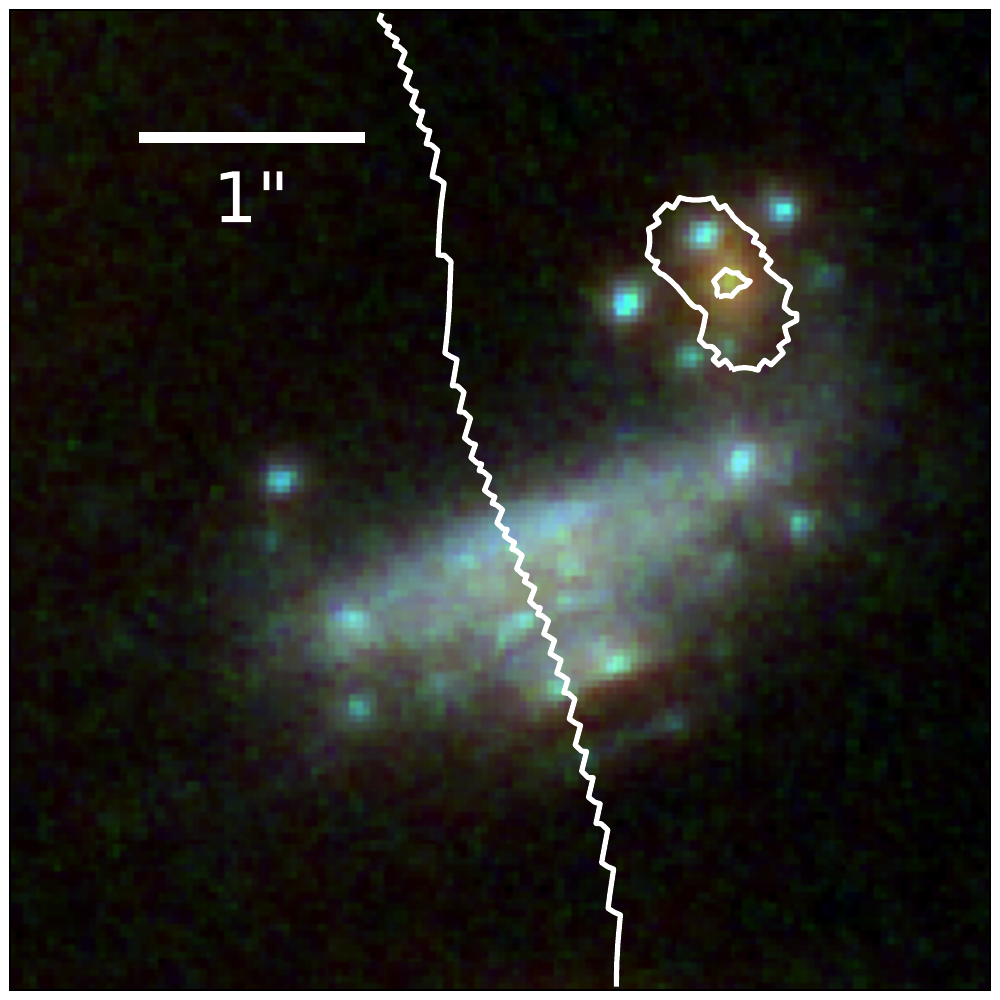}
    \caption{False color RGB images in the field of the galaxy cluster SDSS J1226+2152, assembled from the HST images in F110W, F814W and F606W filters respectively. In the left panel, the lensing critical curves corresponding to a source redshift $z=2.93$ are drawn as the white curves, which pass through the magnified arc of interest. The right panel is a $4.5\arcsec \times 4.5\arcsec$ image cutout centered at (RA, DEC) = (12:26:51.31, +21:52:19.62). The exact location of intersection between the critical curve and the arc is subject to uncertainty in the macro lens model. The arc forms as two images of the source galaxy merge at the position of the critical curve, which results in the mirror-symmetric
    appearance of the arc. In fact, the highly magnified arc constrains the local position and shape of the critical curve better than other faraway lensed sources used to construct the macro model.}
    \label{fig:image}
\end{figure*}
%%%%%%%%%%%%%%%%%%%%%%%%%%%%%%%%%%%%%%%%%%%%%%%%%%%%%%%%%%%%%%%%%%%%

%%%%%%%%%%%%%%%%%%%%%%%%%%%%%%%%%%%%%%%%%%%%%%%%%%%%%%%%%%%%%%%%%%%%
\section{Data}
\label{sec:data}
%%%%%%%%%%%%%%%%%%%%%%%%%%%%%%%%%%%%%%%%%%%%%%%%%%%%%%%%%%%%%%%%%%%%

The caustic-straddling lensed arc SGAS 122651.3+215220 was discovered in December 2007 at the 2.5m Nordic Optical Telescope~\citep{2010ApJ...723L..73K} and is in the catalog of SDSS Giant Arcs Survey (SGAS; \cite{2008AJ....135..664H}). The arc redshift is determined to be $z_s=2.9260\pm0.0002$ from nebular lines~\citep{rigby2018magellan}. The spectroscopic redshift of the galaxy cluster lens SDSS J1226+2152 $z_l=0.43$ was reported by \cite{2014ApJ...783...41B}. Data were used to perform lens modeling with the \texttt{LENSTOOL} software package~\citep{2007NJPh....9..447J}. It approximates the cluster mass distribution as a linear superposition of mass halos and then uses Markov Chain Monte Carlo sampling to determine the best-fit set of parameters for the lens model. The fitting procedure for analogous clusters is described in \cite{2019arXiv190405940S}.

The arc S1226 was imaged in multiple HST filters. We use for our analysis images taken with the Advanced Camera for Surveys (ACS) in two optical filters, F606W and F814W, and with the Wide Field Camera 3 (WFC3) in two IR filters, F110W and F160W. The images in the optical filters were obtained 6 years prior to the images obtained in the IR filters. \reftab{obs} summarizes the observations. In \reffig{image}, we show a composite false-color image of the cluster and a zoomed-in view of the arc.

The arc will be observed with the Near Infrared Spectrograph (NIRSpec) and the Mid-Infrared Instrument (MIRI) integral field unit (IFU) and imaged with the Near Infrared Camera (NIRCam) in multiple filters aboard JWST, as part of the DD-ERS program (see \reftab{obsj} for details). 

%%%%%%%%%%%%%%%%%%%%%%%%%%%%%%%%%%%%%%%%%%%%%%%%%%%%%%%%%%%%%%%%%%%%

\begin{savenotes}
\begin{table*}
	\centering
	\caption{Archival HST imaging of SDSS J1226+2152 used in this work. Multiple visits in individual filters are combined into single drizzled science images sampled at a pixel scale of $0.03\arcsec$.}
	\label{tab:obs}
	\begin{tabular}{cccccc} % four columns, alignment for each
		\hline
		\hline
		Instrument & Filter & Exposure [ks] & Date & Program \# & PI \\
		\hline
		ACS & F606W & 2.0 & 2011-04-14 & 12368 & Morris \\
		ACS & F606W & 2.0 & 2011-04-15 & 12368 & Morris \\
		ACS & F814W & 2.0 & 2011-04-14 & 12368 & Morris \\
		ACS & F814W & 2.0 & 2011-04-15 & 12368 & Morris \\
		WCF3 & F110W & 1.2 & 2017-12-18 & 15378 & Bayliss \\
		WCF3 & F160W & 1.3 & 2017-12-18 & 15378 & Bayliss\\
		\hline
	\end{tabular}
\end{table*}
\end{savenotes}

%%%%%%%%%%%%%%%%%%%%%%%%%%%%%%%%%%%%%%%%%%%%%%%%%%%%%%%%%%%%%%%%%%%%

%%%%%%%%%%%%%%%%%%%%%%%%%%%%%%%%%%%%%%%%%%%%%%%%%%%%%%%%%%%%%%%%%%%%
\begin{table*}
	\centering
	\caption{Forthcoming JWST observations with proposed filters/methods and exposure times (\# 1335, PI: Rigby)~\citep{rigby2017templates}.}
	\label{tab:obsj}
	\begin{tabular}{ccc} % four columns, alignment for each
		\hline
		\hline
		Instrument & Filters/methods & Exposure [ks] \\
		\hline
		NIRSPEC & F170LP, G235H (IFU) & 8.3 \\
		MIRI & F560W, MRS (IFU) & 3.6 \\
		NIRCAM & F115W, F277W & 0.3 \\
		NIRCAM & F150W, F356W & 0.3 \\
		NIRCAM & F200W, F444W & 0.3 \\
		\hline
	\end{tabular}
\end{table*}
%%%%%%%%%%%%%%%%%%%%%%%%%%%%%%%%%%%%%%%%%%%%%%%%%%%%%%%%%%%%%%%%%%%%

%%%%%%%%%%%%%%%%%%%%%%%%%%%%%%%%%%%%%%%%%%%%%%%%%%%%%%%%%%%%%%%%%%%%
\section{Asymmetries in the arc}
\label{sec:asym}
%%%%%%%%%%%%%%%%%%%%%%%%%%%%%%%%%%%%%%%%%%%%%%%%%%%%%%%%%%%%%%%%%%%%

According to our macro lens model, the portion of the critical line that passes through the arc is nearly a straight line and is approximately perpendicular to the direction of arc elongation (shown as the white curve in \reffig{image}), except that a small cluster member galaxy to the northwest (also shown in \reffig{image} with the small critical curve it generates) slightly deforms the surface brightness profile of the arc. The focus of this study is on the brightest part of the arc on the south which is not significantly influenced by the galaxy perturber (also see discussion in \refsec{galperturber}). Images of the arc in four HST filters are shown in \reffig{slits}, after rotation to make the principal axis of elongation of the arc horizontal. Our five slits are shown in the bottom panel, on the same image of the arc as in \reffig{image}. If the lens surface mass distribution is locally smooth, a symmetric appearance across the critical curve is expected. 

In \refsec{slits}, we first qualitatively show evidence for departure from symmetry using a simple method which was previously applied to a similar caustic-straddling arc in MACS J0416.1-2403~ \citep{2019ApJ...880...58K}. In this method, we align multiple slits with the direction of arc elongation, and identify prominent bright features on both sides of the critical curve that are likely to be image pairs of major underlying compact sources. We then undertake a rigorous measurement of their flux inequality that is described in \refsec{fluxasym}.

%%%%%%%%%%%%%%%%%%%%%%%%%%%%%%%%%%%%%%%%%%%%%%%%%%%%%%%%%%%%%%%%%%%%
\subsection{Slits}
\label{sec:slits} 
%%%%%%%%%%%%%%%%%%%%%%%%%%%%%%%%%%%%%%%%%%%%%%%%%%%%%%%%%%%%%%%%%%%%

\begin{figure}
    \centering
    \includegraphics[width=\linewidth]{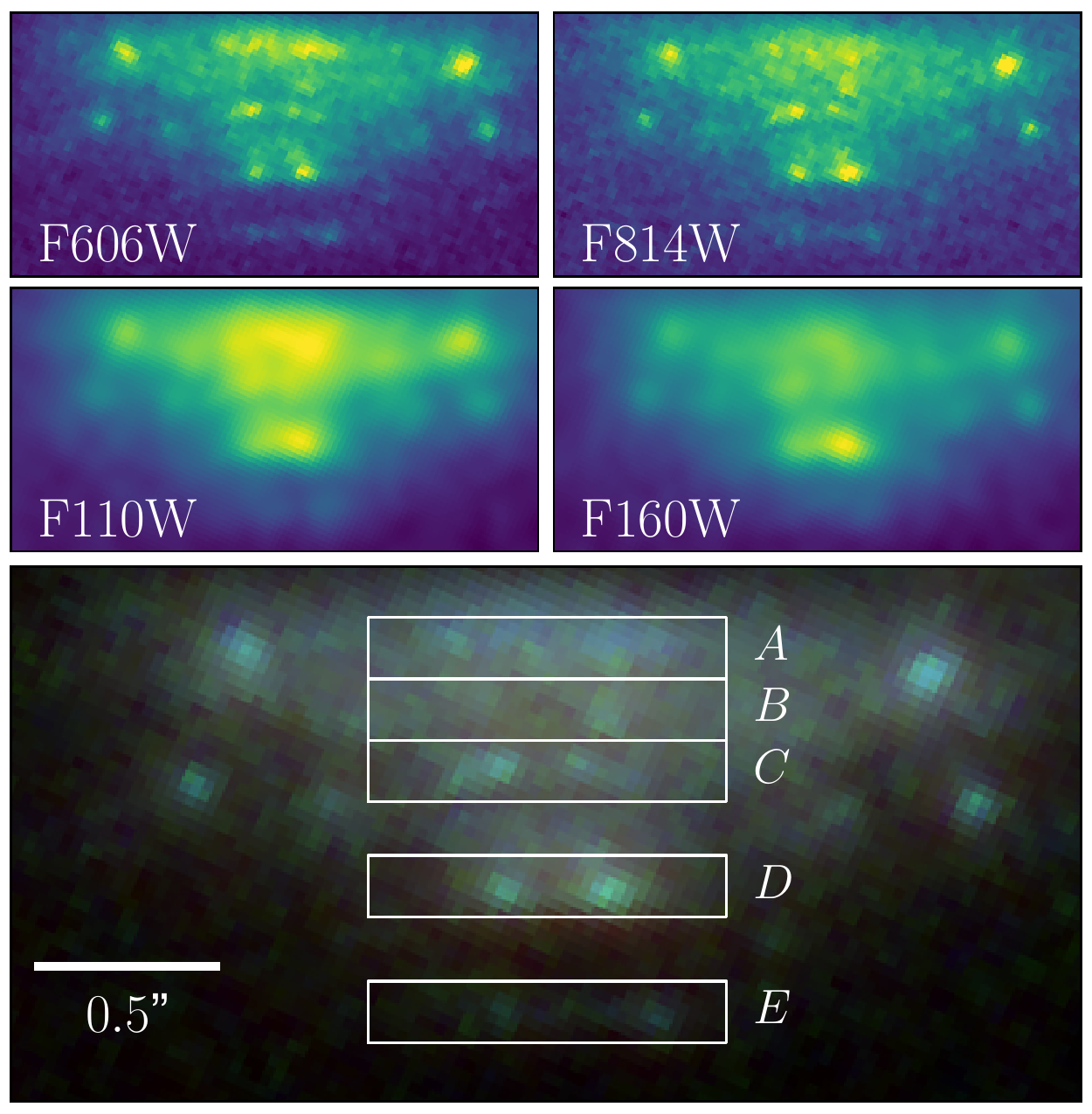}
    \caption{Asymmetric surface brightness features in the arc. Top four panels separately show four HST filters. Bottom panel shows the same false-color image as in \reffig{image} with the five $1\arcsec \times 0.17\arcsec$ slits, defined to be roughly perpendicular to the cluster critical curve. The brightness profiles along these slits are shown in \reffig{profiles}. The cluster critical curve is nearly vertical in this figure, and the slits are parallel to the direction of arc elongation.}
    \label{fig:slits}
\end{figure}

\begin{figure}
    \centering
    \includegraphics[scale=0.9]{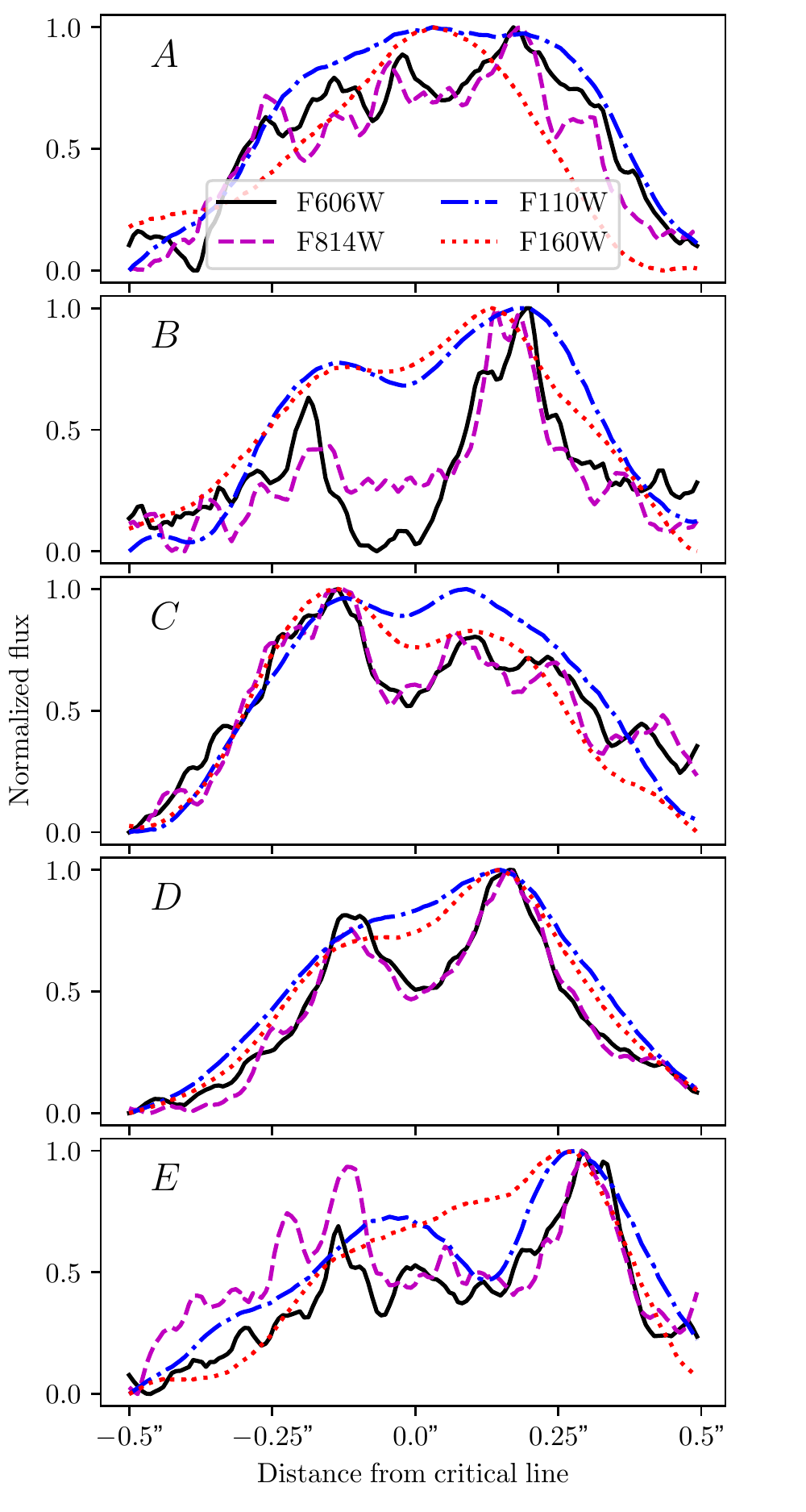}
    \caption{Brightness profiles along several $1\arcsec \times 0.17\arcsec$ slits as defined in \reffig{slits}. Slits B, C and D have a major pair of peaks discernible in all four HST filters. }
    \label{fig:profiles}
\end{figure}

We define five $1\arcsec \times 0.17\arcsec$ slits along the critical curve, each of which is positioned to align with the direction of arc elongation and to include a pair of major surface brightness features, as shown in \reffig{slits}. For every HST filter, we sum the flux along the perpendicular direction, and plot the summed flux as a function of the position along the slit direction in \reffig{profiles}. 

In Slit D, a major pair of flux peaks in both F606W and F814W is clear, with sizable asymmetries that are compatible between the two filters. The pair of peaks is also present in F110W and F160W, but are broadened by the wider point spread functions (PSFs). In each of Slits B, C and E, a major pair of unequal peaks are also visible in F606W and F814W, but other minor peaks make the pair identification less clear. Asymmetric double-peaked profiles in rough correspondence are also visible in the two IR filters. The profiles in slit A are more complex with multiple peaks of comparable heights and are harder to interpret.

To further explore the presence of asymmetric image pairs, we apply in \refapp{nonrig} a non-physical technique to search for asymmetric features which flexibly accounts for small deviations from perfect fold symmetry. As shown there, this method highlights the asymmetric features in Slits C and D in F606W and F814W, as well as the feature in Slit E in F606W. The most outstanding feature in Slit D is also clear from the results for F110W and F160W.

We interpret the major flux double peaks in Slits B, C, D and E as pairs of highly magnified images of luminous, compact star-forming regions near the macro caustic. For these image pairs with sub-arcsecond separation, significantly asymmetric magnifications are however unexpected from a smooth lens. Note that other image pairs at larger separations, $\gtrsim 1\arcsec$, may have unequal magnifications that can be explained by higher order corrections to an idealized fold lens model near the critical curve.

%%%%%%%%%%%%%%%%%%%%%%%%%%%%%%%%%%%%%%%%%%%%%%%%%%%%%%%%%%%%%%%%%%%%
%%%%%%%%%%%%%%%%%%%%%%%%%%%%%%%%%%%%%%%%%%%%%%%%%%%%%%%%%%%%%%%%%%%%

%%%%%%%%%%%%%%%%%%%%%%%%%%%%%%%%%%%%%%%%%%%%%%%%%%%%%%%%%%%%%%%%%%%%
%%%%%%%%%%%%%%%%%%%%%%%%%%%%%%%%%%%%%%%%%%%%%%%%%%%%%%%%%%%%%%%%%%%%

%%%%%%%%%%%%%%%%%%%%%%%%%%%%%%%%%%%%%%%%%%%%%%%%%%%%%%%%%%%%%%%%%%%%
\subsection{Flux asymmetries}
\label{sec:fluxasym} 
%%%%%%%%%%%%%%%%%%%%%%%%%%%%%%%%%%%%%%%%%%%%%%%%%%%%%%%%%%%%%%%%%%%%

Motivated by the findings of \refsec{slits} through eyeballing slit flux profiles, we now set out to measure asymmetric fluxes carefully. We relegate extra technical details of this analysis to \refapp{analysisdetails}.

Adopting the assumption that lensed images are underresolved, we model major macro lensed images with appropriate PSFs. This assumption is valid if sources are individual stars or star groups that are more compact than $\lesssim 2\,{\rm pc}\,(100/\mu)$, where $\mu$ is the magnification.

In each filter separately, and for each asymmetric feature under study, we first create a suitable cutout which consists of $N_{\rm pix}$ pixels and contains the presumptive image pair. The cutout is chosen to be large enough to capture most information provided by the PSFs of the image pair, but cropped to exclude complex neighboring surface brightness structures. We evaluate a likelihood function $\mathcal{L}=\exp(-\chi^2/2)$, where we define the chi-squared function:
\begin{align}
    \label{eq:chisq}
    \chi^2 = \sum^{N_{\rm pix}}_{i=1}\,\frac{1}{\sigma^2_i}\left[ d_i - \sum^{2}_{j=1}\,F^{(j)}_\nu\,p^{(j)}_i(\bfx^{(j)}) - \sum^{n_{\rm bkg}}_{j=1}\,c^{(j)}\,b^{(j)}_i(\bfp^{(j)}) \right]^2.
\end{align}
Here $d_i$ is the flux in pixel $i$, which we fit to the superposition of two PSFs representing a pair of unresolved images centered at $\bfx^{(j)}$ with fluxes $F^{(j)}_\nu$ for $j=1,\,2$ respectively, and $n_{\rm bkg}$ components of diffuse surface brightness background with profiles $b^{(j)}_i$, parameterized by an additional set of parameters $\bfp^{(j)}$, for $j=1,2,\cdots,n_{\rm bkg}$, respectively. We introduce the inverse weights $\sigma_i$ to quantify flux errors, making the simplifying assumption that errors are uncorrelated among pixels. In this work, we set all $\sigma_i$'s to be a uniform number $\sigma_F$ (i.e. the white noise approximation). Refer to \refapp{analysisdetails} for our choices for the diffuse background profiles $b^{(j)}_i$ and $\sigma_F$.

In \refeq{chisq}, the noise $\sigma_i$ should not be interpreted as a photometric error only. In fact, the HST images are deep enough that detector noise and photon shot noise are subdominant. We adopt the philosophy that $\sigma_F$ mainly reflects our ignorance about the details of an underlying population of minor surface brightness structures on the arc,
acting as an effective noise background when we fit the image pairs. To avoid overfitting due to over-constrained background templates used in \refeq{chisq}, we set an empirical value for $\sigma_F$ by requiring that the $\chi^2$ per effective degree of freedom in \refeq{chisq}, for the maximal likelihood solution, should be around unity. In this way, we include modeling uncertainties about fainter surface brightness structures into our quoted errorbars.

%%%%%%%%%%%%%%%%%%%%%%%%%%%%%%%%%%%%%%%%%%%%%%%%%%%%%%%%%%%%%%%%%%%%
\begin{figure*}
    \centering
    \includegraphics[scale=0.4]{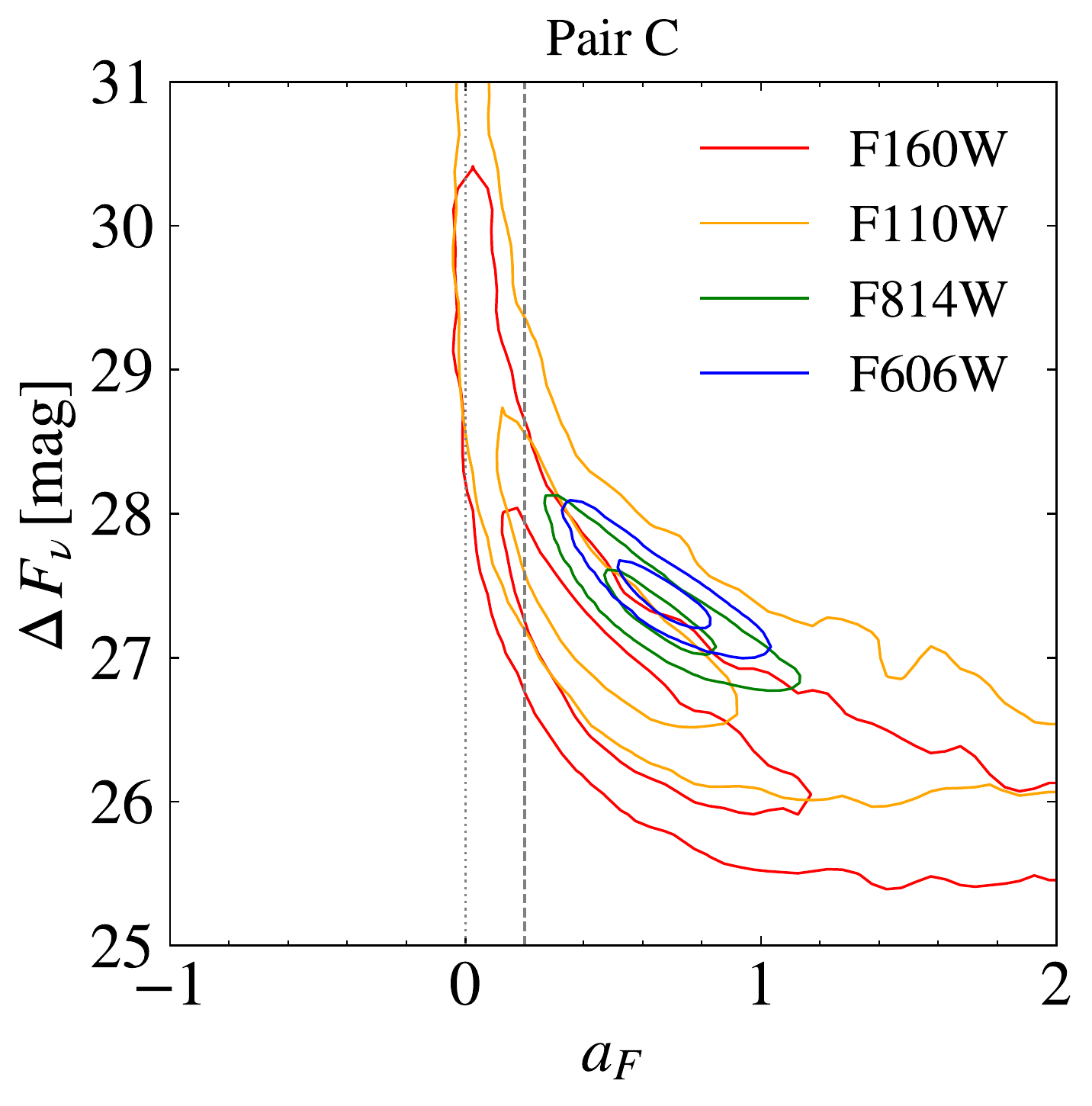}
    \includegraphics[scale=0.4]{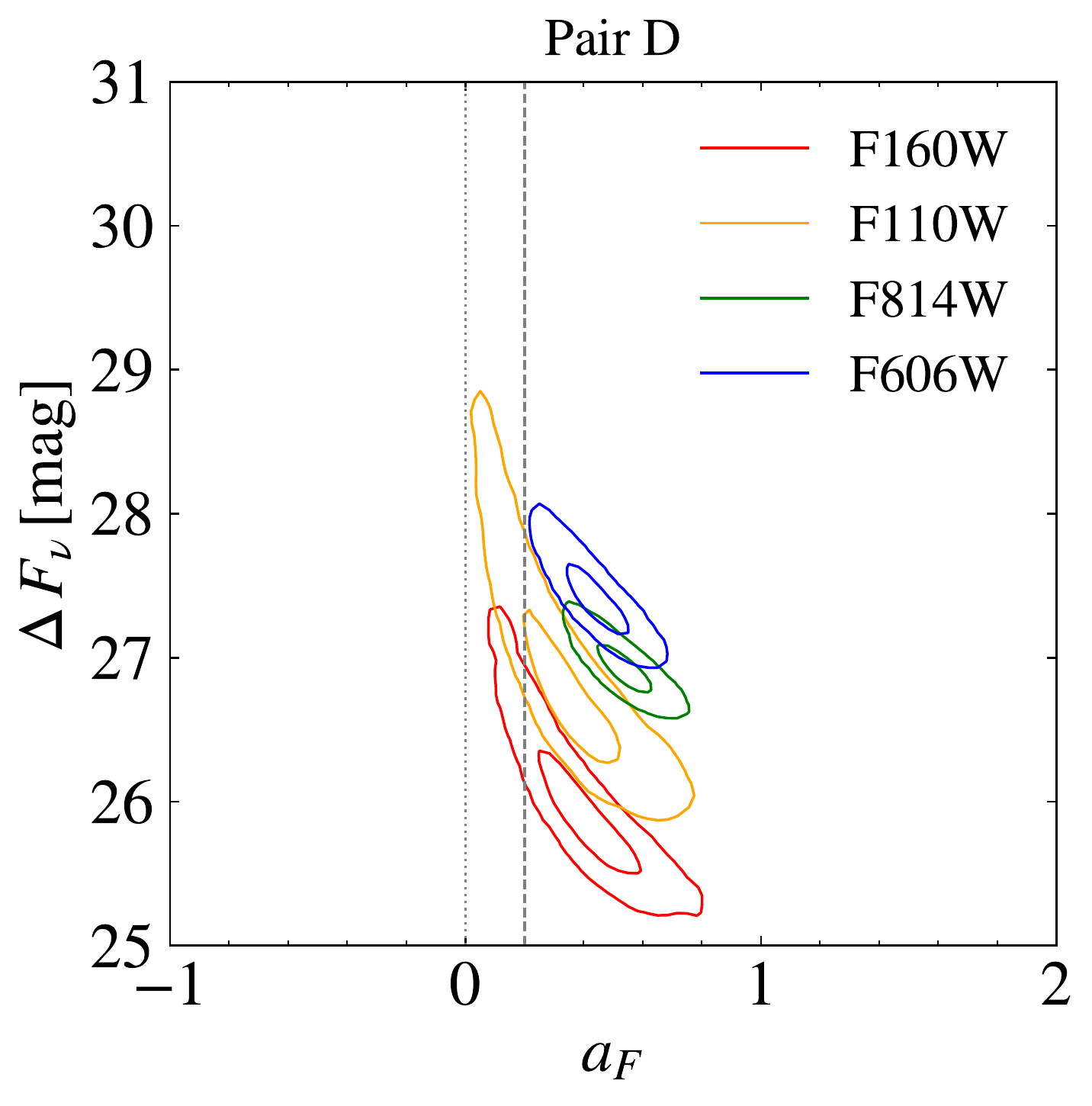}
    \includegraphics[scale=0.4]{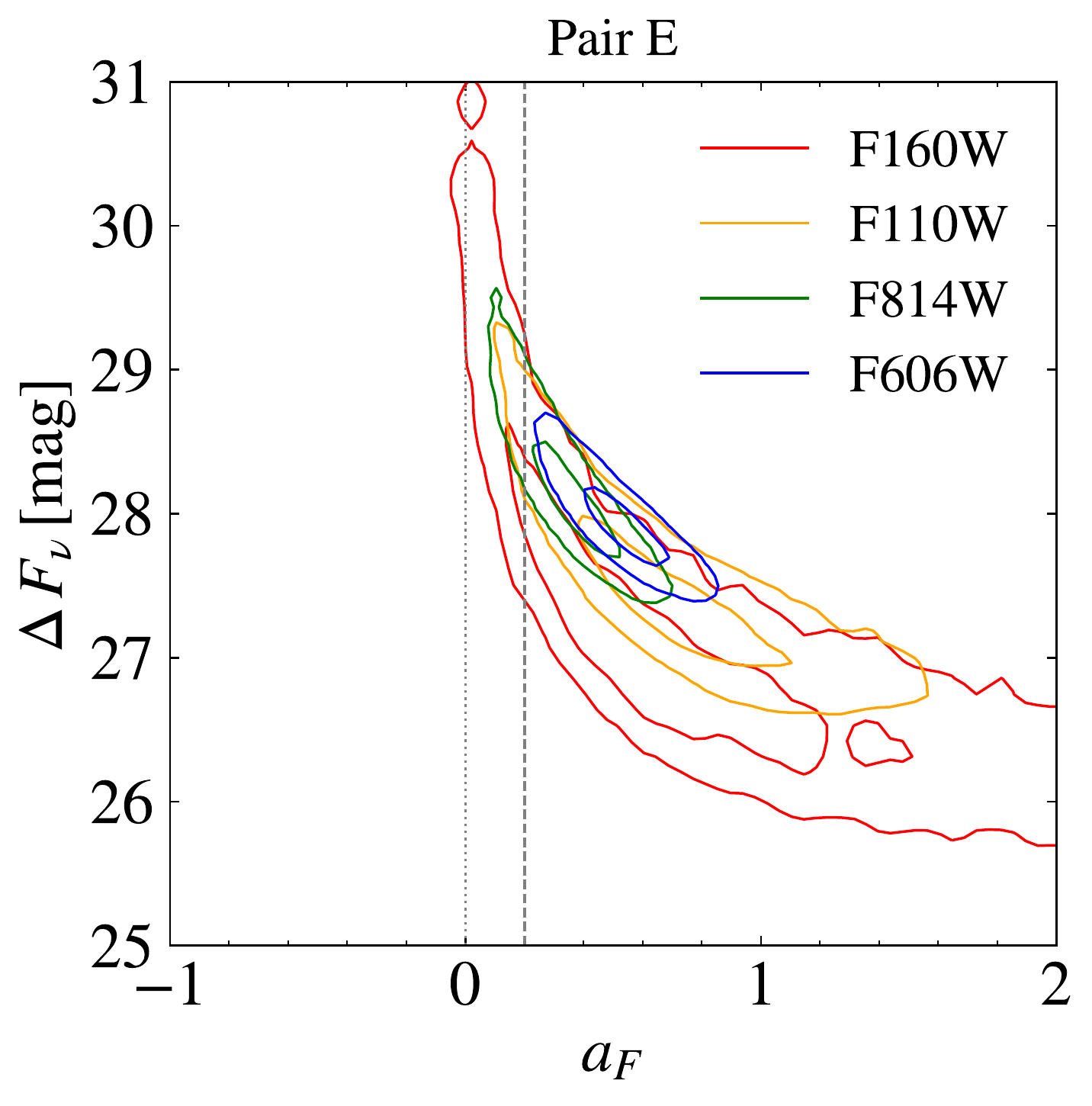}
    \caption{Joint posterior distributions for the fractional flux asymmetry $a_F$ and flux difference $\Delta F_\nu$, marginalized over all other fitting parameters, for the three asymmetric image pairs we identify in Slits C, D and E, and for the four HST filters. Inner and outer contours enclose 50\% and 95\% of the posterior distribution, respectively. The dashed (dotted) vertical line marks $a_F=0.2$ ($a_F=0$).
    }
    \label{fig:slits_CDE_asy}
\end{figure*}
%%%%%%%%%%%%%%%%%%%%%%%%%%%%%%%%%%%%%%%%%%%%%%%%%%%%%%%%%%%%%%%%%%%%

The likelihood evaluation is then coupled to the sampler \texttt{PyMultiNest}~\citep{Buchner:2014nha} to derive the joint posterior distribution of the model parameters. Flat priors are adopted for all parameters over conservatively wide sampling ranges.

We apply the above methodology to the asymmetric image pairs in Slit C, D and E, which are visually discernible in all four HST filters. From the measured normalizations for the two fitted PSFs, we compute the fractional flux asymmetry
\begin{align}
    \label{eq:aF}
a_F = 2\,\frac{F^{(1)}_{\nu} - F^{(2)}_{\nu}}{F^{(1)}_{\nu} + F^{(2)}_{\nu}}~,
\end{align}
in accord with the fractional magnification asymmetry that will be defined later in \refeq{amu}, and the absolute flux difference $\Delta F_\nu = F^{(1)}_{\nu} - F^{(2)}_{\nu}$, where $F^{(1)}_{\nu}$ ($F^{(2)}_{\nu}$) is the flux of the brighter (dimmer) image of the pair. In \reffig{slits_CDE_asy}, we show the measured values for $a_F$ and $\Delta F_\nu$ with their uncertainties.

The image pair in Slit D has a most probable value $a_F = 0.4$--$0.6$ from the F606W and F814W filters, with a robust lower bound $a_F > 0.2$. The results from the F110W and F160W filters are fully consistent with $a_F = 0.4$--$0.6$ but have larger uncertainties. The flux inferred for either of the lensed image pair monotonically increases from the bluest filter to the reddest one, consistent with the color of young, hot stars or young stellar association at $z_s\sim 3$. In particular, the pair flux difference $\Delta F_\nu$ is safely brighter than 28th magnitude in F606W and F814W. These results are consistent with lensed macro image pairs with significantly different but achromatic magnification factors, persistent over a timescale of about six years.

Flux asymmetries of features in Slits C and E are also robustly detected in the F606W and F814W filters, with probable values in the range $a_F = 0.2$--$1.0$. Asymmetries and absolute flux measurements in F110W and F160W are subject to large uncertainties because centroiding of point sources is affected by blending in these two filters with wide PSFs. The asymmetry is nevertheless compatible with that seen in F606W and F814W, and a source color consistent with star-forming regions. We therefore find evidence that even the features in Slits C and E exhibit significant asymmetry in optical filters, although consistency in IR filters remains unproven.

Having demonstrated that asymmetric image pairs arise in several places along the critical curve, with at least one bright pair (Slit D) being well measured, in the next Sections we will explore the physical interpretation of this phenomenon.

%%%%%%%%%%%%%%%%%%%%%%%%%%%%%%%%%%%%%%%%%%%%%%%%%%%%%%%%%%%%%%%%%%%%
\section{Macro lens model near critical curve}
\label{sec:macrolens}
%%%%%%%%%%%%%%%%%%%%%%%%%%%%%%%%%%%%%%%%%%%%%%%%%%%%%%%%%%%%%%%%%%%%

At the location of arc, we infer from the cluster lens model a ratio $\kappa_0 = \Sigma_0/\Sigma_{\rm crit} = 0.8$ between the coarse-grained total surface mass density $\Sigma_0$ of the lens at the line of sight and the critical surface mass density~\citep{blandford1986fermat} $\Sigma_{\rm crit}=(c^2/4\pi\,G)\,(D_S/D_L\,D_{LS})\approx 2\times 10^9\,M_\odot/{\rm kpc}^2$, where $D_L$, $D_S$ and $D_{LS}$ are the angular diameter distances to the cluster lens at $z_l=0.43$, to the source galaxy at $z_s=2.93$, and from the lens to the source. This is the local coarse-grained value for the lensing convergence of the macro lens. Near the critical curve, a compact source appears as a pair of macro images, each of which has an unsigned magnification factor
\begin{align}
    \label{eq:barmu}
    \bar{\mu} = \frac{1}{2\,|\Delta\theta|\,|1-\kappa_0|\,d\,|\sin\alpha|}
\end{align}
where $\Delta\theta$ (signed) equals to half the angular separation between the image pair, $d$ is the gradient of inverse magnification in the critical curve vicinity, and $\alpha$ is the angle between the critical curve and the direction of arc elongation. Numerically, we measure from our cluster lens model $d \approx 7.5\,{\rm arcmin}^{-1}$ and $\alpha \approx 90^\circ$, which gives $\bar{\mu} \approx 140\,\left(|\Delta \theta|/0.14\,\arcsec\right)^{-1}$. While the lensed arc considered here at $z_s=2.92$ is substantially more distant than the caustic straddling lensed galaxies studied in MACS J1149.5+2223~\citep{2018NatAs...2..334K} and MACS J0416.1-2403, the magnification factor as a function of image separation across the critical curve is similar in order of magnitude to the values found in those systems.

%%%%%%%%%%%%%%%%%%%%%%%%%%%%%%%%%%%%%%%%%%%%%%%%%%%%%%%%%%%%%%%%%%%%
\section{Intracluster stars}
\label{sec:icstars}
%%%%%%%%%%%%%%%%%%%%%%%%%%%%%%%%%%%%%%%%%%%%%%%%%%%%%%%%%%%%%%%%%%%%

The line of sight to the arc is at a projected distance of $B \sim 50\,{\rm kpc}$ to the BCG. Intervening intracluster stars can originate from several sources: (i) diffuse light extending from the BCG~\citep{zwicky1951coma, lin2004k, zibetti2005intergalactic}; (ii) diffuse light extending from a major cluster member galaxy $\sim 4 \arcsec$ south of the arc; (iii) diffuse light extending from a minor cluster member galaxy just $\sim 2\arcsec$ away from the arc to the northwest. The minor member galaxy acts as a substructure lens, causing a compact and bright star-forming source to appear in multiple images, as can be seen in the right panel of \reffig{image}. The colors measured from the 4 HST filters are consistent between components (i) and (ii), while it is difficult to reliably determine the colors of component (iii) due to arc contamination. Inferred from the image taken in the reddest filter F160W, component (iii) can be safely neglected, and component (ii) is subdominant compared to the diffuse light halo of the BCG, if not entirely negligible.

The major axis of the BCG diffuse light nearly intersects the lensed arc. We fit the isophotes in each HST filter to ellipses centered at the BCG and adopt the same ellipticity for all four filters. We find consistent colors wherever contamination from other sources are negligible. The surface brightness at the location of the arc calculated from the fitted isophote ellipses is within a factor of two when compared to a few nearby places that are not on top of the arc. 

We use the population synthesis code Flexible Stellar Population Synthesis (\texttt{FSPS}) to model intracluster stars~\citep{conroy2009propagation, conroy2010propagation} assuming the initial mass function of \cite{kroupa2001variation}. We find that the colors are consistent with an old simple stellar population, with a single age and metallicity. The age ranges from $t_{\rm ssp} = 7\,$Gyr for low metallicity $\log(Z/Z_\odot) = -1.0$ to $ t_{\rm ssp} = 1.5\,$Gyr for high metallicity $\log(Z/Z_\odot) = 0.3$. Varying $\log(Z/Z_\odot)$ from $-1.0$ to $0.3$ correspond to a conservative range $0.002 < \kappa_\star < 0.009$. While breaking the age-metallicity degeneracy necessitates spectroscopic analysis, we think the most likely age is $t_{\rm ssp} \approx 3\,$Gyr with metallicity $\log(Z/Z_\odot) \simeq -0.3$, a value found typical for intracluster light (ICL) at similar cluster-centric radii in many cluster lenses of similar redshifts~\citep{Montes:2017yct}. This translates into a local surface density of intracluster stars $\Sigma_{\rm ICL} \approx 10^7\,M_\odot/{\rm kpc}^2$ and hence a convergence $\kappa_\star = \Sigma_{\rm ICL}/\Sigma_{\rm crit} \approx 0.005$, which we use hereafter as our fiducial value. This value is comparable to previous estimates carried out for caustic straddling lensed galaxies in MACS J1149.5+2223~\citep{PhysRevD.97.023518} and in MACS J0416.1-2403~\citep{2019ApJ...880...58K}.

In calculating $\kappa_\star$, we have included into the synthesized stellar population white dwarfs in addition to main sequences and evolved stars, but have neglected neutron stars and black holes. The latter group are in any case remnants of massive stars $M \gtrsim 8\,M_\odot$, which account for only a sub-dominant fraction $\sim 14\%$ of the initial star-forming mass. In our model, intracluster stars have masses in the range $0.08$--$1.4\,M_\odot$. Insensitive to modeling uncertainty, the most abundant microlenses are main sequence dwarfs whose masses are around $\sim 0.2\,M_\odot$.

%%%%%%%%%%%%%%%%%%%%%%%%%%%%%%%%%%%%%%%%%%%%%%%%%%%%%%%%%%%%%%%%%%%%
\section{Intracluster Microlensing}
\label{sec:microlensing}
%%%%%%%%%%%%%%%%%%%%%%%%%%%%%%%%%%%%%%%%%%%%%%%%%%%%%%%%%%%%%%%%%%%%

The observed flux of a compact source is affected by stellar microlenses near the line of sight, which render the two macro images unequally bright at a given epoch. One possibility is that each observed asymmetric pair of images reflects the microlensing phenomenon acting on either a {\it single} super-luminous source star or a compact {\it group} of bright source stars in the lensed galaxy. Under this hypothesis, no substructure in the lens mass distribution other than that in individual stellar microlenses needs to be invoked to explain the asymmetries. 

Intracluster stars should break the smooth macro critical curve into an interconnected network of micro critical curves, whose full width is $2\,\kappa_\star/d = 0.03$--$0.14\arcsec$, with our best estimate being $0.08\arcsec$. This network band has one edge at the expected location of the macro critical curve, and the other edge lies on its interior~\citep{2017ApJ...850...49V}. Under the assumption that the macro critical curve roughly bisects the asymmetric pairs of surface brightness features in \reffig{slits}, none of the features lie within this band. This implies that microlensing variability is infrequent and occurs over a long timescale for any single lensed star.

%%%%%%%%%%%%%%%%%%%%%%%%%%%%%%%%%%%%%%%%%%%%%%%%%%%%%%%%%%%%%%%%%%%%
\subsection{Microlensing of a single source star}
%%%%%%%%%%%%%%%%%%%%%%%%%%%%%%%%%%%%%%%%%%%%%%%%%%%%%%%%%%%%%%%%%%%%

\begin{figure*}
    \centering
    \includegraphics[scale=0.4]{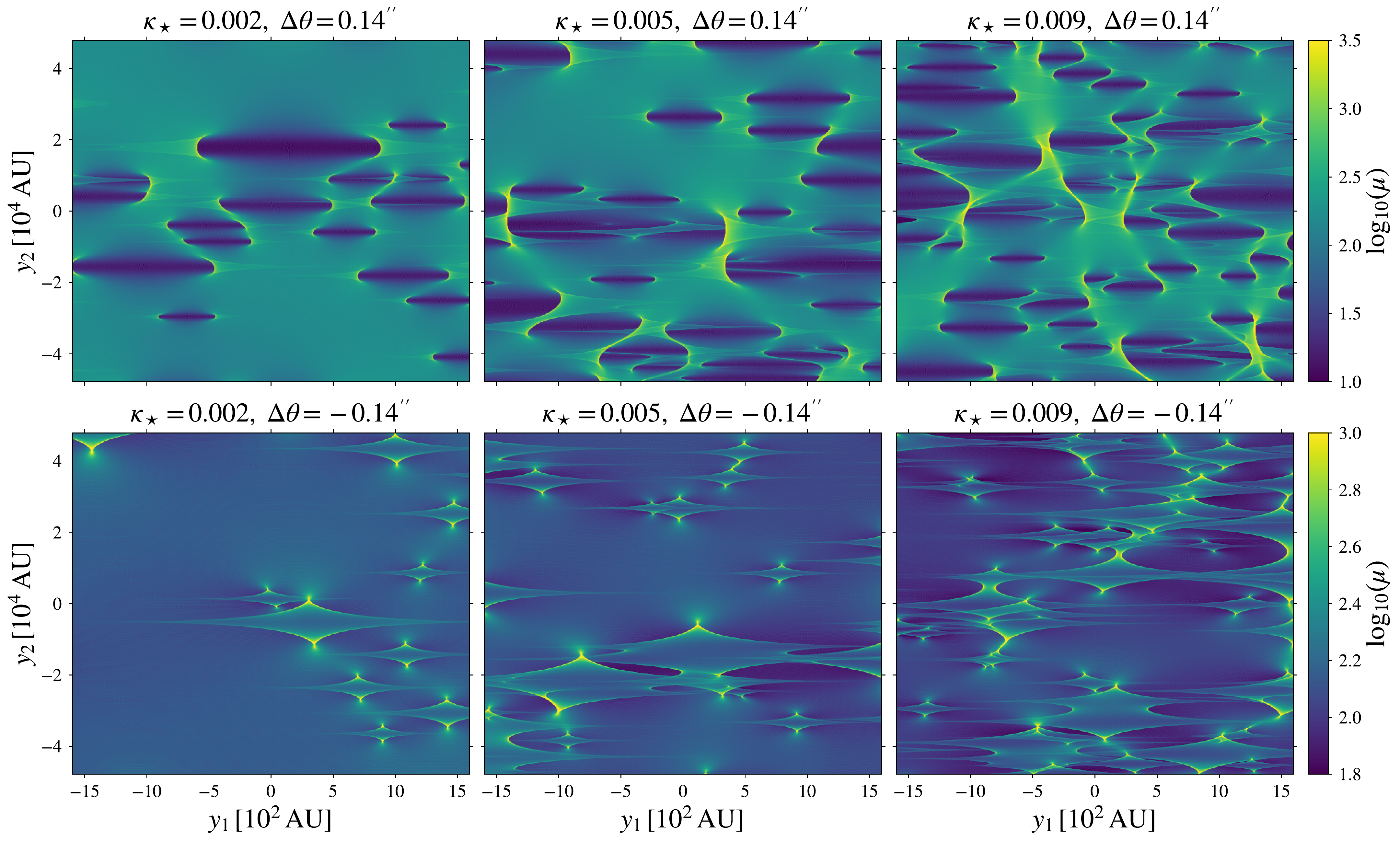}
    \caption{
    Magnification maps of random source-plane realizations for one of the two macro images of a point source appropriate for the caustic straddling arc S1226. The source-plane region in each panel maps to the image-plane vicinity of a macro image that is $\Delta\theta=\pm 0.14\arcsec$ away (appropriate for image pairs in Slits C and D) from the smooth cluster critical curve. We do not consider DM substructure lensing in these plots. The macro lens model near the intersection of the arc and the cluster critical curve is assumed to be a simple fold model~\citep{1992grle.book.....S} with parameters $\kappa_0 = 0.8$, $d=7.5\,{\rm arcmin}^{-1}$ and $\alpha=90^\circ$. The three columns correspond to three likely values for the total convergence from intracluster microlenses: $\kappa_\star=0.002$, $0.005$, and $0.009$. The top (bottom) row corresponds to the macro image on the interior (exterior) of the smooth critical curve with $\Delta\theta = 0.14\arcsec$ ($\Delta\theta = -0.14\arcsec$). The coordinate system is oriented such that $y_1$-axis and $y_2$-axis map to the two eigen-directions of the lensing Jacobian matrix of the macro lens model on the image plane, with $y_1$-axis corresponding to the direction of nearly vanishing eigenvalue. Scales along the $y_1$ and $y_2$ axes are highly different for better visualization.}
    \label{fig:mumaps}
\end{figure*}

To corroborate this, in \reffig{mumaps} we simulate random realizations of the total microlensing magnification factor for {\it one} of the macro images across a small region on the source plane, assuming a macro lens model appropriate for S1226, and stellar microlens masses randomly drawn from a Simple Stellar Population model with age $t_{\rm ssp}=3\,$Gyr and $\log(Z/Z_\odot)=-0.3$. The resolution of these magnification maps is $\sim 1\,{\rm AU}$. For $\kappa_\star=0.002$--$0.009$ and $\Delta\theta=\pm 0.14\arcsec$, appropriate e.g. for the prominent asymmetries in Slits D and C, the range of source-plane length scales indicates that the magnification may vary by order unity on a timescale of several years for a typical effective source-lens relative velocity $v_t$ (see Eq.(12) of \cite{2017ApJ...850...49V}) of $\sim 100\, {\rm AU/yr}$.

%%%%%%%%%%%%%%%%%%%%%%%%%%%%%%%%%%%%%%%%%%%%%%%%%%%%%%%%%%%%%%%%%%%%
%%%%%%%%%%%%%%%%%%%%%%%%%%%%%%%%%%%%%%%%%%%%%%%%%%%%%%%%%%%%%%%%%%%%

If the asymmetric fluxes observed in several slits (e.g. Slits C, D and E, and especially Slit D) are due to microlensing of a {\it single} source star, the pairs of macro images must have large differences in their magnification factors. At $z_s = 2.93$, only the most luminous source stars in the rest-frame UV under unusually high magnification can explain the observed flux asymmetries reported in \reffig{slits_CDE_asy}. 
The best constraints are obtained from the F814W filter because the asymmetric features are brighter than in F606W, and sharper than in F110W and F160W. In F814W, the image pair flux difference is safely brighter than magnitude $m = 28.5$ in Slits C and D, and is likely as bright as $m = 28.5$ in Slit E. Assuming $m_{\Delta F_{\nu, {\rm F814W}}} \leqslant 28.5\,$mag, which is conservative for Slits C and D, we list in \reftab{srcstarmag} the required minimum magnification differences, $|\Delta\mu_{\rm min}|$, needed to explain the measured flux asymmetries for several candidate types of stars.

The best candidate source stars are low-metallicity blue supergiants (BSG), but even these require a magnification asymmetry $|\Delta\mu| \gtrsim 600$. The most luminous main-sequence (MS) stars require much higher magnifications of $|\Delta\mu| \gtrsim 3000$, which are much less likely. This is under the optimistic assumption of zero dust reddening, while \cite{2019ApJ...882..182C} report E(B-V)$=0.13\,$mag measured for S1226. Since we have conservatively compute for $m_{\Delta F_{\nu, {\rm F814W}}} \leqslant 28.5\,$mag, the magnification difference needs to be another factor of four larger in order to explain the asymmetry in Slit D.

There is no evidence for rare stars (e.g. zero metallicity, extremely massive stars) that might be much brighter than those listed in \reftab{srcstarmag}. The rest-frame UV spectrum of the arc analyzed by \cite{2019ApJ...882..182C} suggests that S1226 has a moderately sub-solar metallicity $-0.9 <\log(Z/Z_\odot)<-0.4$ and a stellar age $\sim 20$--$26\,$Myr, and is dominated by B-type giants with no trace of O stars detected. The best stellar candidates for the image pairs that can produce asymmetries through microlensing are therefore BSGs as listed in \reftab{srcstarmag}.    

\begin{figure}
    \centering
    \includegraphics[scale=0.56]{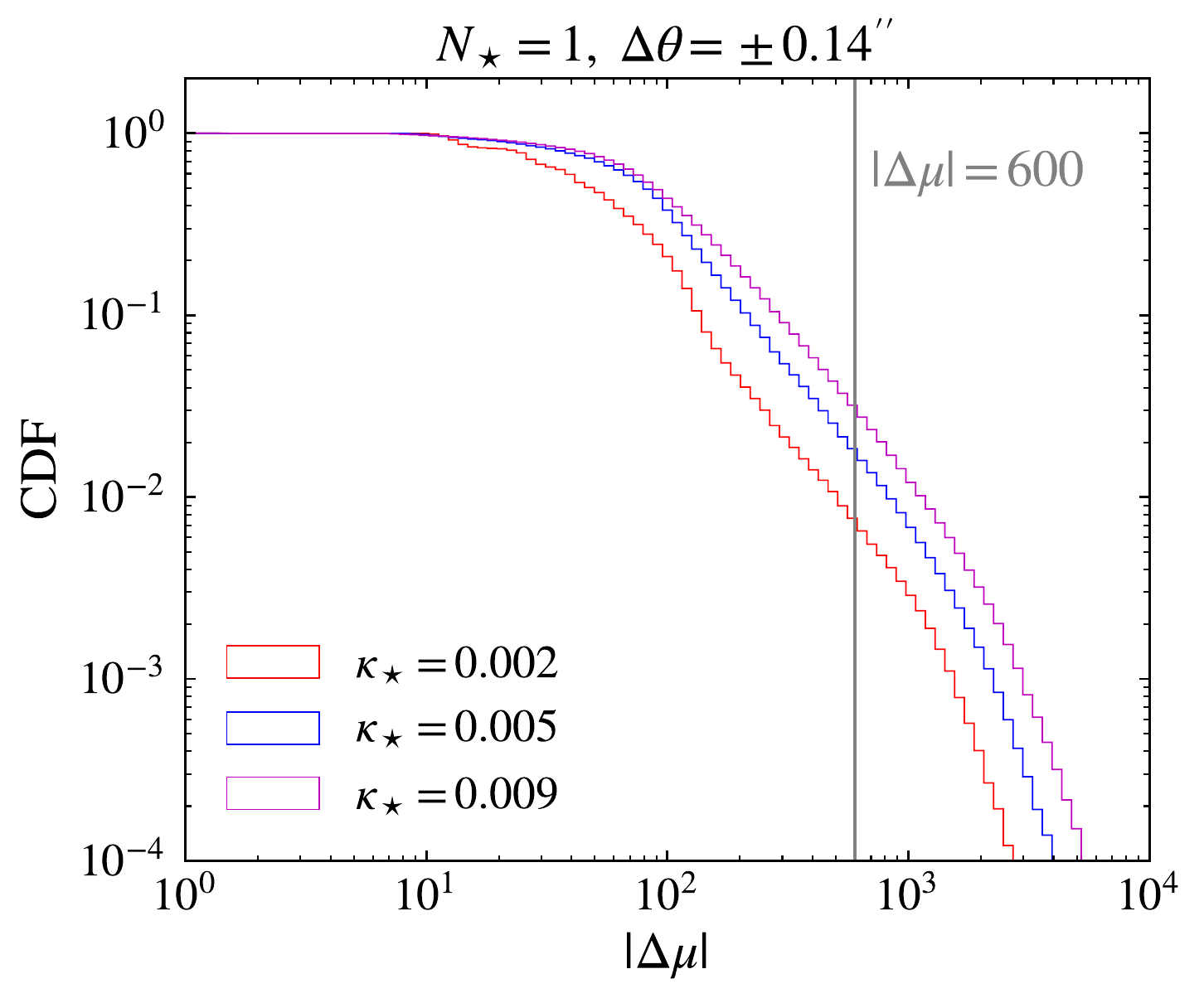}
    \caption{Numerically derived cumulative distribution for the magnification difference $\Delta\mu$ between the two macro images of a single source star. The assumption of a point source is valid for either MS or giant star across the range of $|\Delta\mu|$ shown here. Curves are plotted for $\kappa_\star=0.002$, $0.005$, and $0.009$. The grey vertical line marks the required minimum magnification asymmetry $|\Delta\mu| \simeq 600$ for the example BSGs considered in \reftab{srcstarmag}. The tail of the distribution at $|\Delta\mu| \gtrsim {\rm few} \times 10^3$ artificially steepens due to finite pixels in numerical ray shooting. The choice $\Delta\theta=\pm 0.14\arcsec$ is appropriate for Slits C and D. The grey vertical line indicates the required magnification difference for the most luminous blue supergiants available in our model stellar population, and suggests that a single source star is unlikely to account for the observed flux asymmetry.}
    \label{fig:mu_CDF_N1}
\end{figure}

The required values of $|\Delta\mu|$ in \reftab{srcstarmag} clearly exceed the expected macro magnification $\bar{\mu} \approx 140$ by a large factor, and are rarely reached as seen in \reffig{mumaps}. 
This large magnification difference is only possible when the source star undergoes a micro caustic transit. \reffig{mu_CDF_N1} indicates that $|\Delta\mu| \gtrsim 600$ randomly occurs for less than a few percent of the time. Microlensing of a single source star is therefore unlikely to be the cause of the flux asymmetries seen in the arc. The main difference between the case of S1226 and the caustic straddling galaxies lensed by MACS J1149.5+2223 and MACS J0416.1-2403, which have confirmed microlensed individual stars, is the lower redshift of the latter sources, $z=1.49$ and $0.94$, making it much more likely for a microlensed luminous star to reach the faintest observable fluxes.

%%%%%%%%%%%%%%%%%%%%%%%%%%%%%%%%%%%%%%%%%%%%%%%%%%%%%%%%%%%%%%%%%%%%
\begin{table*}
	\centering
	\caption{Examples of ultra-luminous stars and their properties as generated by \texttt{FSPS}: (1) stellar age $t_{\rm age}$; (2) metallicity $\log(Z/Z_\odot)$; (3) stellar mass $M_\star$; (4) effective temperature $T_{\rm eff}$; (5) stellar radius $R_\star$; (6) bolometric luminosity $L_{\rm bol}$; (7) minimum magnification difference $|\Delta\mu|_{\rm min}$ for an image pair flux asymmetry $m_{\rm F814W}=28.5$ as seen from $z_s = 2.93$ in the absence of dust reddening (conservative lower bound for the measured flux asymmetry of image pairs in Slits C and D; see \reffig{slits_CDE_asy}). The flux asymmetry measured in F814W for the image pair in Slit D is larger by another magnitude, requiring even higher magnifications. Redshifted from $z_s=2.93$, only very massive hot O-type MS stars or evolved BSGs are sufficiently bright in the F814W filter; smaller or colder stellar types would require implausibly large $|\Delta\mu|_{\rm min} \gtrsim 10^4$. Even for hot O-type MSs or BSGs, $|\Delta\mu|_{\rm min}$ is much greater than the predicted macro magnification factor in \refeq{barmu}.}
	\label{tab:srcstarmag}
	\begin{tabular}{cccccccc}
		\hline
		\hline
		stellar type & $t_{\rm age}\,[{\rm Myr}]$ & $\log\left(Z/Z_\odot\right)$  & $M_\star\,[M_\odot]$ & $T_{\rm eff}\,[{\rm K}]$ & $R_\star\,[R_\odot]$ & $L_{\rm bol}\,[10^6\,L_\odot]$ & $|\Delta\mu|_{\rm min}$ \\
		\hline
		blue supergiant & 3 & $0.0$ & $42$ & $22000$ & $64$ & $0.88$ & $1869$ \\
		blue supergiant & 3 & $-0.3$ & $47$ & $13000$ & $274$ & $1.9$ & $1064$ \\
		blue supergiant & 3 & $-1.0$ & $66$ & $16000$ & $205$ & $2.4$ & $655$ \\
		blue supergiant & 3 & $-2.0$ & $66$ & $16000$ & $200$ & $2.4$ & $645$ \\
		main sequence & 0.3 & $0.0$ & $115$ & $50000$ & $17.5$ & $1.7$ & $3317$ \\
		main sequence & 0.3 & $-0.3$ & $117$ & $52000$ & $16$ & $1.7$ & $3565$ \\
		main sequence & 1 & $-1.0$ & $107$ & $53000$ & $15$ & $1.6$ & $4127$ \\
		main sequence & 0.3 & $-2.0$ & $118$ & $60000$ & $12$ & $1.7$ & $4085$ \\
		\hline
	\end{tabular}
\end{table*}
%%%%%%%%%%%%%%%%%%%%%%%%%%%%%%%%%%%%%%%%%%%%%%%%%%%%%%%%%%%%%%%%%%%%

%%%%%%%%%%%%%%%%%%%%%%%%%%%%%%%%%%%%%%%%%%%%%%%%%%%%%%%%%%%%%%%%%%%%
%%%%%%%%%%%%%%%%%%%%%%%%%%%%%%%%%%%%%%%%%%%%%%%%%%%%%%%%%%%%%%%%%%%%

%%%%%%%%%%%%%%%%%%%%%%%%%%%%%%%%%%%%%%%%%%%%%%%%%%%%%%%%%%%%%%%%%%%%
\begin{figure}
    \centering
    \includegraphics[scale=0.56]{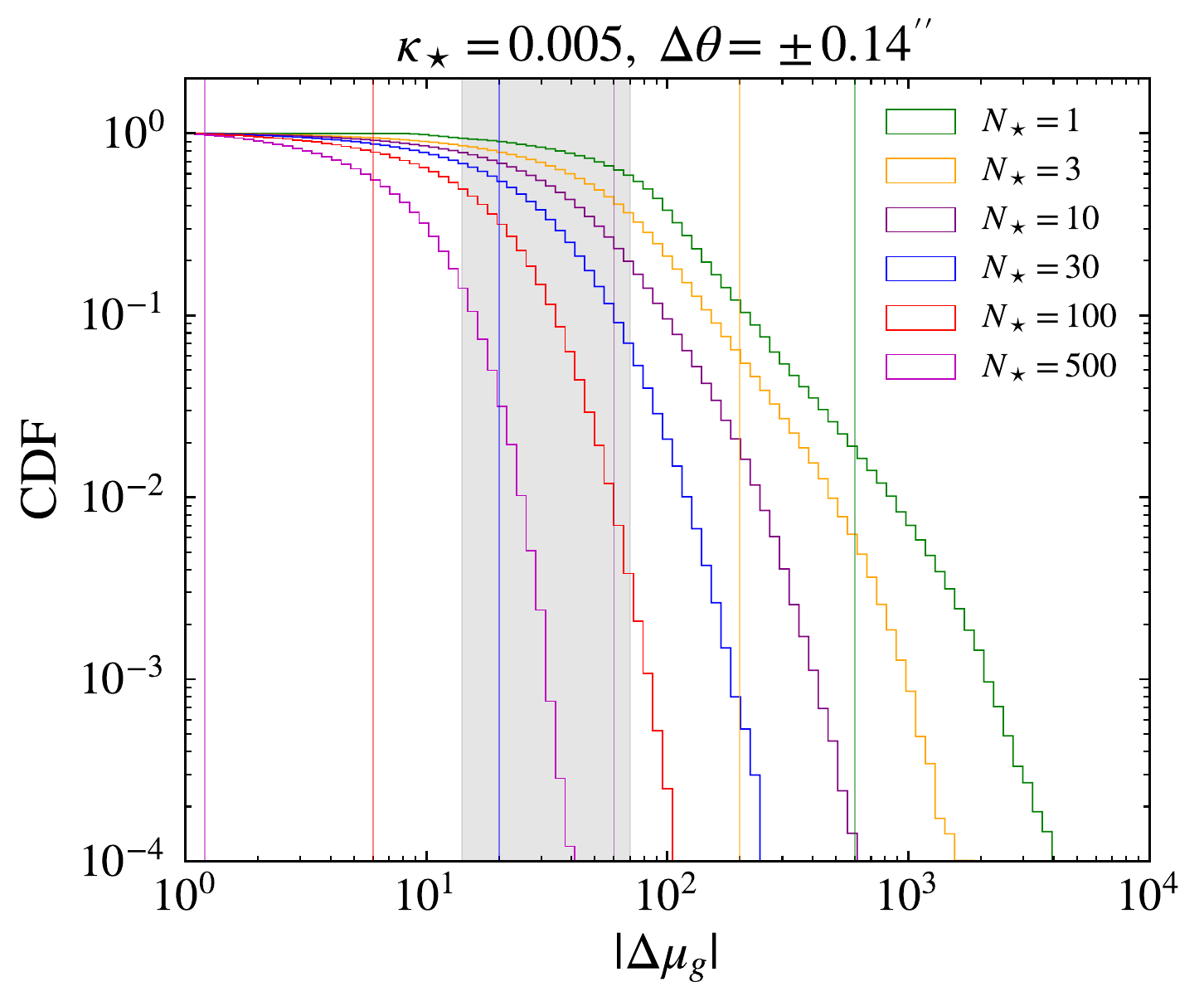}
    \caption{Cumulative distribution for the magnification asymmetry $\Delta\mu_g$ for $\kappa_\star=0.005$. This is similar to \reffig{mu_CDF_N1} but for a group of $N_\star$ identical source stars subject to uncorrelated microlensing. The grey shaded band corresponds to $10$\%--$50$\% magnification asymmetry estimated for the asymmetric features shown in \reffig{profiles}. For each value of $N_\star$ (color coded), a vertical line is drawn to indicate $|\Delta\mu_g|=600/N_\star$. The location of the grey shaded band in this plot favors $N_\star\sim 10$--$30$ brightest stars.}
    \label{fig:mu_CDF_multi_stars}
\end{figure}
%%%%%%%%%%%%%%%%%%%%%%%%%%%%%%%%%%%%%%%%%%%%%%%%%%%%%%%%%%%%%%%%%%%%

%%%%%%%%%%%%%%%%%%%%%%%%%%%%%%%%%%%%%%%%%%%%%%%%%%%%%%%%%%%%%%%%%%%%
\subsection{Microlensing of a group of source stars}
%%%%%%%%%%%%%%%%%%%%%%%%%%%%%%%%%%%%%%%%%%%%%%%%%%%%%%%%%%%%%%%%%%%%

The large asymmetric fluxes are more naturally explained by a group of luminous stars (possibly co-existing with many more fainter stars). Indeed, hot luminous stars commonly cluster in compact star-forming regions or open clusters. The age of the arc stellar population, estimated to be $\sim 20$--$26\,$Myr from rest-frame UV spectroscopy~\citep{2019ApJ...882..182C}, is longer than that of the BSGs in \reftab{srcstarmag}, which implies that the stellar light is dominated by giant B stars that are individually less massive and luminous than the BSGs in \reftab{srcstarmag}.

The asymmetric image pairs, e.g. in Slits C and D, appear barely resolved in the F814W image. Given the angular distance from the cluster critical curve, our macro lens model constrains their physical sizes to be $\lesssim 5\,$pc. Neglecting closely bound multiple stars for the time being, member stars within the group should have mutual separations on the order of parsecs, which is much greater than the source-plane correlation lengths $\sim 10^2$--$10^3\,{\rm AU}$ of microlensing magnifications as shown in \reffig{mumaps}. 

This implies that at any given time individual luminous stars have statistically independent microlensing magnifications. Under this assumption, we show in \reffig{mu_CDF_multi_stars} the probability distribution of $\Delta\mu_g$ for a star group, where $\mu_g$ is the (microlensing affected) magnification averaged over $N_\star$ stars for one (unresolved) {\it macro} image of the star group on one side of the macro critical curve. As $N_\star$ increases, the distribution of $\Delta\mu_g$ approaches a Gaussian centered at zero with decreasing width. For simplicity, we assume that all $N_\star$ stars are identical. A small number $N_\star$ is disfavored for the same reason as in the previous subsection (typical values of $|\Delta\mu_g|$ are insufficient to explain the flux asymmetries). The minimum $|\Delta\mu_g|$ to account for the flux asymmetries is reduced as $\propto 1/N_\star$. We find that a value $N_\star \sim 10$--$30$ is most consistent with the level of macro image asymmetry seen in \reffig{profiles}. This number yields a typical magnification difference $|\Delta\mu_g|$ induced by uncorrelated microlensing of $\sim 10\%$--$50\%$ of $\bar{\mu}\approx 140$, shown as the shaded vertical band in \reffig{mu_CDF_multi_stars}. A number $N_\star \sim 100$--$500$ of stars of comparable brightness might still produce a ratio $|\Delta\mu_g|/\bar{\mu}$ compatible with \reffig{mu_CDF_multi_stars}, but each star should then be fainter in order not to overproduce the observed flux of either macro image. For $N_\star \gtrsim 500$, $\Delta\mu_g$ would be highly diluted and a flux asymmetry as large as $\sim 10\%$--$50\%$ would be implausible. For a more accurate analysis, one could replace the simplifying assumption of identical source stars with a model of a continuous luminosity function. We leave such analysis for future investigation.

If $N_\star$ ultra-luminous stars form a system more compact than $\sim 10^2\,$AU, their microlensing magnifications will be highly correlated. They may simultaneously have a magnification asymmetry $|\Delta\mu| \simeq 600/N_\star$. For $N_\star \sim 3$--$6$, the value for $|\Delta\mu_g| \approx |\Delta\mu|$ would be typical of the distribution shown in \reffig{mu_CDF_N1}. The lower required magnification may then be more likely to occur, but the tight star clusters required may not be abundant enough to explain the multiple observed asymmetric image pairs.

To summarize, by measuring unequal image pairs at just one random epoch per HST filter, we conclude that while the hypothesis of microlensing of a single ultra-luminous source star is hardly viable, the asymmetries might be explained by a group of stars, under one of the following three situations: uniform microlensing of $N_\star \sim 3$--$6$ extremely luminous BSGs tightly bound within $\lesssim 10^2\,$AU, or uncorrelated microlensing of a group of $N_\star \sim 10$--$30$ such BSGs within a region of size $\lesssim \mathcal{O}({\rm pc})$, or uncorrelated microlensing of a cluster of $N_\star \sim 100$--$500$ less luminous stars (most likely B-type giants) within a region of size $\lesssim \mathcal{O}({\rm pc})$. As we will see next, frequent temporal variability of the flux asymmetry is generally expected from these microlensing scenarios.

%%%%%%%%%%%%%%%%%%%%%%%%%%%%%%%%%%%%%%%%%%%%%%%%%%%%%%%%%%%%%%%%%%%%
\subsection{Temporal variability of flux asymmetry}
%%%%%%%%%%%%%%%%%%%%%%%%%%%%%%%%%%%%%%%%%%%%%%%%%%%%%%%%%%%%%%%%%%%%

The flux asymmetry between the two macro images induced by microlensing is expected to vary with time as each source star slowly traverses the microlensing magnification pattern on the source plane (as in \reffig{mumaps}). As a result, the asymmetry can frequently change its sign.

\begin{figure*}
    \centering
    \includegraphics[scale=0.42]{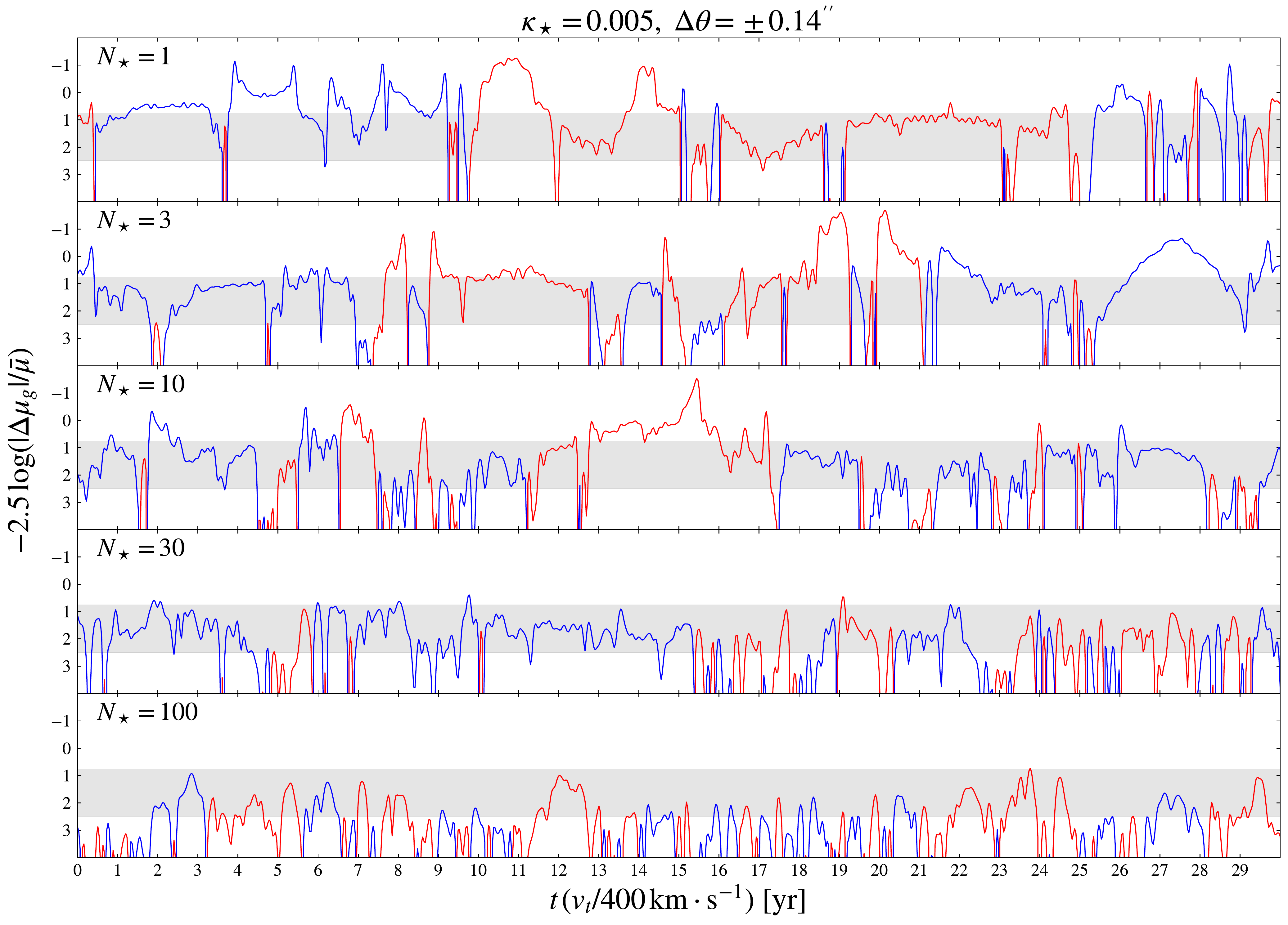}
    \caption{Numerical random realizations of the microlensing induced temporal variability for the fractional magnification asymmetry $\Delta\mu_g/\bar{\mu}$ between the two macro images over a baseline of decades. The panels correspond to different values of $N_\star$, the number of identical source stars. The timescales indicated on the horizon axis can be rescaled with the effective source-lens relative transverse velocity $v_t$ along the direction of high elongation, for which we use a fiducial value $v_t = 400\,{\rm km/s}$. The vertical axis shows the logarithm $-2.5\,\log\left(|\Delta\mu_g|/\bar{\mu}\right)$, which measures change in the flux asymmetry in units of photometric magnitude. In each light curve, the red (blue) portion corresponds to the macro image to the interior ($\Delta\theta>0$) of the cluster critical curve being brighter (fainter) than the macro image to the exterior ($\Delta\theta<0$) of the cluster critical curve. The horizontal shaded band in grey corresponds to fractional flux asymmetry $|\Delta\mu_g|/\bar{\mu}=10\%$--$50\%$. We note that oscillations of tiny amplitude on very short timescales are pixelization artifacts in our ray-shooting simulation (most visible in the $N_\star=1$ curve).}
    \label{fig:lc_examples}
\end{figure*}

As the star number $N_\star$ of a group increases, the overall flux asymmetry should vary more rapidly because the flux asymmetry from each member star varies independently and there are more frequent micro caustic crossings, but the fractional size of the asymmetry is more diluted. This is clearly reflected in \reffig{lc_examples}, where we show numerical examples of flux asymmetry variation from randomly drawn microlens realizations, up to the uncertainty in the timescale that scales inversely with the unknown effective transverse velocity parameter $v_t$. 

Since the arc has not been imaged at more than one epoch in any single filter, we cannot rule out temporal variation in the flux asymmetries with complete certainty. Still, the asymmetries in Slit D seen in F606W and F814W are very similar to those seen in F110W and F160W, if not identical, even though the two optical filters and the two IR filters were used at two different epochs separated by more than 6 years. This seems to hint that in Slit D the observed flux asymmetry is persistent, or the variable component is relatively small.

Assuming a fiducial velocity $v_t \sim 400\,{\rm km/s}$, up to a factor of few uncertainty unless fine-tuned, the apparent non-variability over $\sim 6\,$yr can be compatible with $N_\star\sim 1$--$3$, but we have seen that such few number of dominant bright stars are probably incompatible with the large size of the asymmetries in absolute flux units. On the other hand, \reffig{lc_examples} suggests that $N_\star \gtrsim 10$ would lead to changes in the sign and magnitude of $\Delta\mu_g$ that occur too frequently over the $\sim 6\,$yr timescale, unless the relative transverse velocity is surprisingly low, $v_t \lesssim 100\,{\rm km/s}$. Extremely tight systems of $N_\star \simeq 3$--$6$ supergiant stars with nearly uniform microlensing might still explain some of the asymmetries, but taking into account all previous considerations, we believe the observed persistent flux asymmetry in Slit D is not in good agreement with microlensing models of either a single or an under-resolved group of source stars. Ultimately, imaging in the same filters at additional epochs is necessary to robustly establish or falsify microlensing variability over a timescale of years.

%%%%%%%%%%%%%%%%%%%%%%%%%%%%%%%%%%%%%%%%%%%%%%%%%%%%%%%%%%%%%%%%%%%%
%%%%%%%%%%%%%%%%%%%%%%%%%%%%%%%%%%%%%%%%%%%%%%%%%%%%%%%%%%%%%%%%%%%%

%%%%%%%%%%%%%%%%%%%%%%%%%%%%%%%%%%%%%%%%%%%%%%%%%%%%%%%%%%%%%%%%%%%%
\section{Substructure lensing}
\label{sec:subhalolensing}
%%%%%%%%%%%%%%%%%%%%%%%%%%%%%%%%%%%%%%%%%%%%%%%%%%%%%%%%%%%%%%%%%%%%

In \refsec{microlensing}, we have studied flux asymmetries from intracluster microlensing operating on minuscule angular scales far beyond telescope resolution. In this Section, we consider substructure lensing from either small galaxies or star-free DM subhalos inside the cluster halo, which also cause flux asymmetries as we show below. The characteristic lensing angular scales of $\sim 0.01$ to $0.1\,{\rm arcsec}$ are marginally resolved in diffraction limited exposures. Unlike microlensing, asymmetric patterns resultant from substructure lensing do not exhibit noticeable variability over timescales of decades to centuries.

%%%%%%%%%%%%%%%%%%%%%%%%%%%%%%%%%%%%%%%%%%%%%%%%%%%%%%%%%%%%%%%%%%%%
\subsection{Galaxy perturbers}
\label{sec:galperturber}
%%%%%%%%%%%%%%%%%%%%%%%%%%%%%%%%%%%%%%%%%%%%%%%%%%%%%%%%%%%%%%%%%%%%

The smooth lens model inevitably deviates from the idealized fold model~\citep{1992grle.book.....S} away from the macro critical curve. Large perturber lenses far away in projection can induce curvature in the macro critical curve on arc second scales. Indeed, the right panel of \reffig{image} shows that our smooth lens model predicts such an example of curvature due to a minor foreground galaxy $\sim 2\arcsec$ to the northwest. However, these large and distant perturbers are usually unable to generate magnification asymmetries larger than $\sim 10\%$ between image pairs with small separation $\Delta\theta$.

Consider an image of a point source on one side of the critical curve, with magnification $\mu$, and its counter image with magnification $\mu'$. Define the signed fractional magnification asymmetry
\begin{align}
\label{eq:amu}
    a_\mu \equiv 2\,\left( \left|\mu \right| - \left|\mu'\right| \right) /\left( \left|\mu \right| + \left|\mu' \right| \right).
\end{align}
Let $M_p$ be the mass of a faraway perturber, $\theta_p$ its Einstein angular scale, and $b$ the angular impact parameter from the image pair. We estimate that $|a_\mu|$ is of the order (derived in \refapp{magasym})
\begin{align}
\label{eq:amuest}
    \left|a_\mu\right| & \sim \left( \theta^2_p\,\Delta\theta \right)/\left( d\,b^4 \right) \nonumber\\
    & \sim 0.04\,\left( \frac{M_p}{10^{11}\,M_\odot} \right)\,\left( \frac{7.5\,{\rm arcmin}^{-1}}{d} \right)\,\left( \frac{2\,\arcsec}{b} \right)^{4}\,\left( \frac{\Delta\theta}{0.14\,\arcsec}\right),
\end{align}
where we assume $b \gg \Delta\theta$ and that the perturber is at the cluster redshift $z_l=0.43$. The perturber galaxy would therefore have to enclose a mass $\sim 10^{12}\,M_\odot$ within $\lesssim 10\,{\rm kpc}$ to produce tens of percent asymmetry. In fact, a blue compact source, probably a young star cluster associated with the arc at $z_s=2.93$, is lensed by this foreground galaxy into multiple images separated by a critical curve loop of size $\sim 1\,\arcsec$. Taking into account that a macro magnification $\sim 30$ should be acting to enhance the effect of the perturber, we find the small size of the critical curve loop inconsistent with the central part of the foreground galaxy (where star light is detected) enclosing $\gtrsim 10^{11}\,M_\odot$.

\begin{figure}
    \centering
    \includegraphics[scale=0.4]{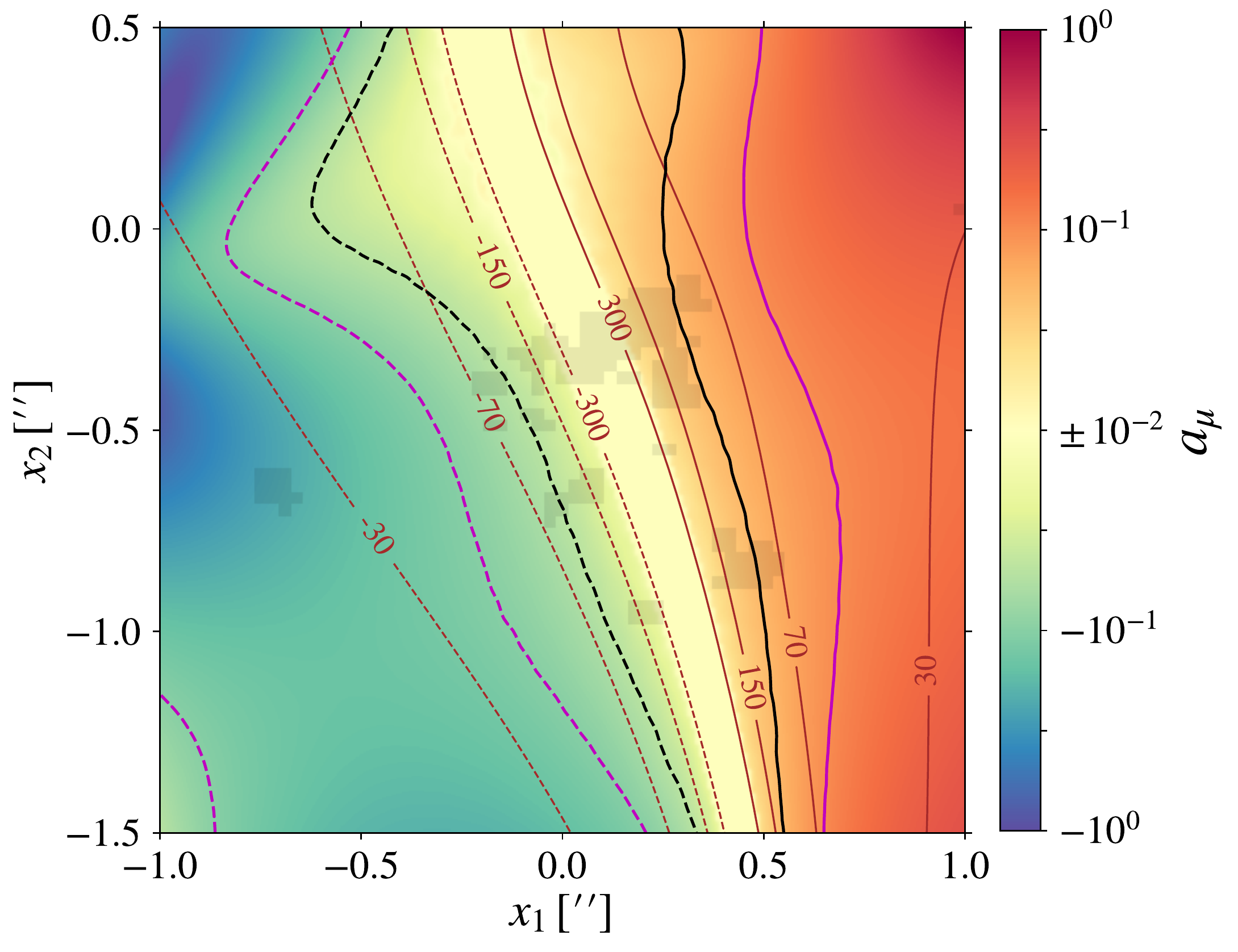}
    \caption{Fractional magnification asymmetry $a_\mu$ (\refeq{amu}) between any image point and its counter image point, according to our smooth lens model and within a $2\arcsec\times 2\arcsec$ FoV centered at the caustic straddling arc and oriented in the same way as \reffig{image}. Black and magenta contours, respectively, correspond to $a_\mu=\pm 0.05$ and $\pm 0.1$, with solid ones for $a_\mu>0$ and dashed ones for $a_\mu<0$. Brown (solid and dashed) curves are contours of constant signed macro magnification factor. The region having $|a_\mu|<10^{-2}$ in the proximity of the critical curve is artificially set to have $a_\mu=\pm 10^{-2}$. For visual guidance, shaded pixels indicate bright features in the F814W filter. We have artificially applied a small, uniform shift of the F814W image relative to the lens model such that the smooth critical curve crudely bisects pairs of surface brightness features. Large magnification asymmetries $|a_\mu|\gtrsim 10\%$ are not expected from a smooth macro lens within $\lesssim 0.2\arcsec$ of the critical curve.}
    \label{fig:mag_asym}
\end{figure}

In \reffig{mag_asym}, we further confirm that the galaxy perturber alone could not have caused the magnification asymmetry. For every image point, we locate the counter image point and calculate $a_\mu$ as defined in \refeq{amu}. In our smooth lens model, $a_\mu$ only reaches a few percent within $\sim 0.2\,\arcsec$ of the critical curve, which is compatible with our order of magnitude estimate in \refeq{amuest}.

%%%%%%%%%%%%%%%%%%%%%%%%%%%%%%%%%%%%%%%%%%%%%%%%%%%%%%%%%%%%%%%%%%%%
%%%%%%%%%%%%%%%%%%%%%%%%%%%%%%%%%%%%%%%%%%%%%%%%%%%%%%%%%%%%%%%%%%%%

%%%%%%%%%%%%%%%%%%%%%%%%%%%%%%%%%%%%%%%%%%%%%%%%%%%%%%%%%%%%%%%%%%%%
\subsection{Dark matter subhalos}
%%%%%%%%%%%%%%%%%%%%%%%%%%%%%%%%%%%%%%%%%%%%%%%%%%%%%%%%%%%%%%%%%%%%

While no evidence supports that any minor foreground galaxy significantly breaks the symmetry near the cluster critical curve, we hypothesize that a population of abundant and non-luminous DM subhalos may be the reason. Unlike the larger but rarer perturbers, sub-galactic subhalos are predicted to be numerous enough to be frequently found close to the critical curves, where their perturbing effects are greatly enhanced~\citep{2017ApJ...845..118M, 2018ApJ...867...24D}.

%%%%%%%%%%%%%%%%%%%%%%%%%%%%%%%%%%%%%%%%%%%%%%%%%%%%%%%%%%%%%%%%%%%%
\begin{figure*}
    \centering
    \includegraphics[scale=0.4]{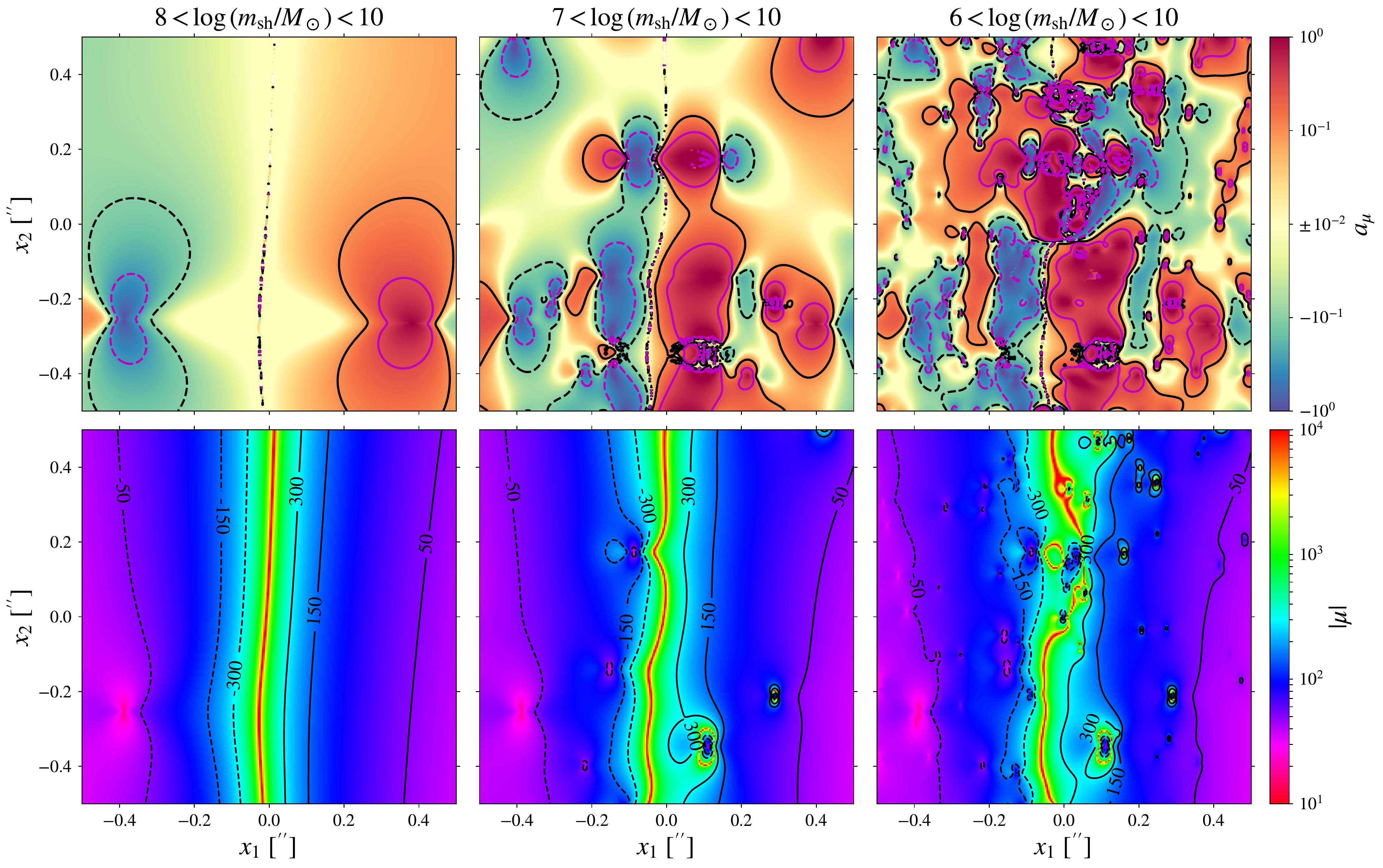}
    \caption{DM subhalo effects on the magnification $\mu$ (bottom panels) and the asymmetry $a_\mu$ (top panels; see \refeq{amu} and the text). A population of $\sim 10^6$--$10^8\,M_\odot$ subhalos render it rare to have magnified image pairs with highly equal fluxes $|a_\mu|\lesssim 1\%$. We zoom into a $1\arcsec \times 1\arcsec$ field, which centers on the cluster critical curve and is oriented such that the $x_1$ axis aligns with the direction of arc elongation. Following the model described in the main text, subhalos less massive than $10^{10}\,M_\odot$ are randomly generated. From left to right, we lower the minimum allowed subhalo mass to $10^8$, $10^7$, and $10^6\,M_\odot$, respectively. Within a projected radius $1\arcsec$ from the coordinate origin, the number of subhalos we generate is $N_{\rm sh}=1$, $27$, and $196$, in the first, second, and third column, respectively. These in total contribute no more than $1\%$ to the local (coarse-grained) surface mass density. In panels of the first column, the only subhalo to the left of the cluster critical curve has a mass $m_{\rm sh}\approx 10^8\,M_\odot$. In the region where subhalos are added, we include a negative uniform convergence to enforce mass conservation. {\it Top panels}: Contours of $a_\mu=\pm 0.1$ ($a_\mu=\pm0.3$) are drawn in black (magenta), with the solid (dashed) line-style indicating $a_\mu>0$ ($a_\mu<0$). Regions having $|a_\mu|<10^{-2}$ are artificially colored according to $a_\mu=\pm 10^{-2}$. Our simulation can have numerical artifacts either when $\mu$ diverges on top of critical curves or when multiple images with very small separations render the solutions to the ray equation inaccurate, but only a negligible fraction of each panel suffers from this problem. {\it Bottom panels}: Contours of constant signed magnification factor $\mu=\pm 50,\,\pm150,\,\pm 300$ are drawn in black, with the solid (dashed) line-style indicating $\mu >0$ ($\mu < 0$).}
    \label{fig:amusubhalo}
\end{figure*}
%%%%%%%%%%%%%%%%%%%%%%%%%%%%%%%%%%%%%%%%%%%%%%%%%%%%%%%%%%%%%%%%%%%%

To numerically assess the effect of subhalos expected from the standard Cold Dark Matter (CDM) theory, we populate the cluster DM halo with randomly generated subhalos, which are superimposed on top of an ideal fold model with parameters appropriate for the arc S1226. We follow the method outlined in \cite{2018ApJ...867...24D} using an extrapolation of the semi-analytic model of \cite{han2016unified} down to subhalo-host mass ratios as small as $m_{\rm sh}/M_{\rm host} \sim 10^{-8}$. For simplicity, we model the host DM halo of SDSS J1226+2152 using a spherical Navarro-Frenk-White (NFW) profile~\citep{1996ApJ...462..563N, navarro1997universal}, with a characteristic mass $M_{\rm 200}=1.5\times 10^{14}\,M_\odot$ and a concentration parameter $C_{\rm 200} \simeq 13$. These broadly reproduce a measured Einstein radius $\theta_E \simeq 10\arcsec$~\citep{2012MNRAS.420.3213O}, a local convergence $\kappa_0=0.8$ at the line of sight as inferred from our macro lens model, as well as an empirical mass-concentration relation found for a catalog of strong and weak lensing SGAS clusters~\citep{bayliss2011gemini, 2012MNRAS.420.3213O}. The host DM halo has an NFW scale radius $R_s = 75\,{\rm kpc}$. Subhalos are assumed to follow a power-law mass function $\rmd n(m_{\rm sh})/\rmd \log m_{\rm sh} \propto m_{\rm sh}^{-0.9}$~\citep{mo2010galaxy}, and are approximated as spherical but tidally truncated NFW profiles~\citep{baltz2009analytic, cyr2016dark}. We refer interested readers to Appendix B of \cite{2018ApJ...867...24D} for technical details.

Similarly to \reffig{mag_asym}, we calculate the magnification asymmetry $a_\mu$ between a pair of images by ray shooting: for every chosen image position, we first find the corresponding source position, and then locate the counter images. When more than one counter image is present, we pick the brightest counter image and evaluate $a_\mu$ according to \refeq{amu}.

In \reffig{amusubhalo}, we show the pattern of $a_\mu$ influenced by one random realization of subhalos for a compact source, within a $1\arcsec \times 1\arcsec$ vicinity of the cluster critical curve. The pattern of signed magnification is also shown. The influence of a subhalo much more massive than $\sim 10^{10}\,M_\odot$ lying on top of the lensed arc should be large enough to be noticeable. But the expected number density of such massive subhalos is low enough that having any one of them within $\sim 1\arcsec$ of the cluster critical curve is improbable. We generate smaller subhalos with $10^6\,M_\odot < m_{\rm sh} < 10^{10}\,M_\odot$. The model of \cite{han2016unified} we adopt here predicts about one subhalo with $m_{\rm sh} \gtrsim 10^8\,M_\odot$ within a disk of $1\arcsec$ radius centered on the cluster critical curve. On the other hand, typically many subhalos with $m_{\rm sh} \sim 10^6\,M_\odot$ are present just a fraction of an arcsecond away from the cluster critical curve. Unable to host star formation, these halos cannot be observed except for gravitational probes sensitive to small scale structures, like the lensing perturbations causing asymmetries in highly magnified image pairs.

As shown in \reffig{amusubhalo}, even one subhalo can induce substantial magnification asymmetry if it is sufficiently close to one of the image pair. A subhalo of mass $m_{\rm sh} \simeq 10^8\,M_\odot$ located $\sim 0.4\arcsec$ from the cluster critical curve can cause $a_\mu \gtrsim 0.1$ within $\sim 0.1\arcsec$--$0.3\arcsec$ from its center. At $\lesssim 0.1\arcsec$ from the cluster critical curve, small subhalos with $m_{\rm sh} \sim 10^6\,M_\odot$ appear common enough to produce $a_\mu \gtrsim 0.1$ across a significant portion of the image plane. Even smaller subhalos should be more numerous and may contribute to sizable $a_\mu$ very close to the critical curve. Our simulation of tiny subhalos $m_{\rm sh} < 10^6\,M_\odot$ is limited by computational cost. Their effects may be important for extremely compact sources such as individual bright stars, but are likely to dilute away for sources of size larger than $\sim 1\,$pc. 

The precise distribution of magnification asymmetry naturally fluctuates among realizations of subhalos, and  mildly depends on host halo structural parameters $M_{\rm 200}$ and $C_{\rm 200}$. The typical magnification asymmetry scales with the normalization of the subhalo mass function. Despite these uncertainties, it seems plausible from our simulations that a population of DM subhalos in the mass range $\sim 10^6$--$10^8\,M_\odot$ induces ubiquitous asymmetry at the level $a_\mu \gtrsim 10\%$ for pairs of highly magnified images within $\sim 0.2\arcsec$ from the cluster critical curve. Subhalos in this mass range are also capable of imprinting astrometric effects detectable with future imaging efforts~\citep{2018ApJ...867...24D}. It is therefore theoretically possible that the flux asymmetries in S1226, such as those in Slits C, D and E, are entirely or partially caused by subhalo lensing. Regarding the ambiguous identification of any image pair in Slit A, for example, we surmise that it could be due to a subhalo creating additional lensed images of the same source. 

%%%%%%%%%%%%%%%%%%%%%%%%%%%%%%%%%%%%%%%%%%%%%%%%%%%%%%%%%%%%%%%%%%%%
\subsection{Satellite galaxies and globular clusters}
\label{sec:}
%%%%%%%%%%%%%%%%%%%%%%%%%%%%%%%%%%%%%%%%%%%%%%%%%%%%%%%%%%%%%%%%%%%%

If sub-galactic perturber lenses are in fact responsible for the observed image pair asymmetries, could DM subhalos be confused by dwarf galaxies in the cluster or along the line of sight, globular clusters or other known stellar systems that are too faint to be detected by HST? Such star clusters or dwarf galaxies would have to be much smaller than, e.g., the minor perturber galaxy to the north of S1226 so as not to induce obvious deformation of the arc, and they would have to be numerous enough to be found near the critical curve with reasonable likelihood. Our ICL estimate at the location of the arc $\kappa_\star \simeq 0.005$ implies that the $1\arcsec\times 1 \arcsec$ FoV in \reffig{amusubhalo} on average only encloses $\sim 10^6\,M_\odot$ of stars, leaving little mass budget to have many stellar mass clumps comparable to the DM subhalos we consider. It is also well known that the general dwarf galaxy luminosity function is much less steep at the faint end than the predicted CDM subhalo mass function at the low-mass end, so dwarf galaxy abundances should be much smaller than required for producing the perturbations illustrated in \reffig{amusubhalo}.

Globular clusters (GCs) populate the intracluster medium due to tidal stripping from their host galaxies~\citep{white1987globular, west1995intracluster}. In recent years, intracluster GCs have been studied in several galaxy clusters up to $z \simeq 0.3$~\citep{2010Sci...328..334L, peng2011hst, west2011globular, Alamo-Martinez:2013aaa, d2016extended, lee2016globular}. In the core of rich clusters, the GC surface number density is found to be around ${\rm few} \times 10^{-2}\,{\rm kpc}^{-2}$~\citep{2010Sci...328..334L, Alamo-Martinez:2013aaa, d2016extended, lee2016globular}, which would correspond to no more than a few GCs within the same FoV in \reffig{amusubhalo}. Even in the extreme case of the core of the Coma Cluster, for which a GC surface density of $\sim 1\,{\rm kpc}^{-2}$ was reported by \cite{peng2011hst}, this number would be $\sim 30$, an order of magnitude smaller than DM subhalos with $m_{\rm sh} \gtrsim 10^6\,M_\odot$. Furthermore, these GCs have masses $\sim 10^5$--$10^6\,M_\odot$ (probably without any significant DM), smaller than those of the DM subhalos we have considered. Hence, interloper GCs are also not abundant enough to be the major cause of the observed image asymmetries.

Satellite galaxies are thought to reside in DM halos. Therefore, satellites should be no more numerous than the DM subhalos that match their host DM halo size, which are already included in the DM subhalo population model we adopt. DM subhalos small enough to be common within the FoV of \reffig{amusubhalo} probably lack any significant stellar component. In the cores of nearby rich galaxy clusters, an extrapolation of the observed satellite luminosity function to the luminosity of GCs~\citep{1995AJ....110.1507B, mobasher2003photometric, 2016ApJ...824...10F} still falls short of the DM subhalo number density seen in \reffig{amusubhalo} by more than two orders of magnitude. Averaging over the entire cluster, we suggest that intracluster satellite galaxies and GCs are less abundant than $m_{\rm sh} \gtrsim 10^6$ DM subhalos to ubiquitously perturb the proximity of the smooth critical curve and induce magnification asymmetries. In the case of S1226, this abundance may be enhanced by satellite galaxies and GCs bound to nearby cluster member galaxies in projection. Deep imaging of the nearby ICL with JWST will set stringent limits on the abundance of faint satellites. Spectroscopic analysis will help rule out foreground stellar systems superimposed on top of the arc.
                    
%%%%%%%%%%%%%%%%%%%%%%%%%%%%%%%%%%%%%%%%%%%%%%%%%%%%%%%%%%%%%%%%%%%%
\section{Conclusions}
\label{sec:disc}
%%%%%%%%%%%%%%%%%%%%%%%%%%%%%%%%%%%%%%%%%%%%%%%%%%%%%%%%%%%%%%%%%%%%

We have investigated HST data showing asymmetric surface brightness features within $\lesssim 0.3\arcsec$ from the lensing critical curve in a $z_s=2.93$ magnified arc behind the galaxy cluster SDSS J1226+2152. We have identified highly magnified image pairs of compact star-forming regions, many of which are inconsistent with being perfectly symmetric. Careful measurement of the brightest of these image pairs robustly constrains the flux ratio to be $>1.2$ in two optical filters, and indicates a most probable flux ratio as large as $\sim 1.3$--$1.8$ in all four HST filters.

Intracluster microlensing inevitably introduces uncorrelated magnification fluctuations in different images of the same source, causing unequal fluxes at any given epoch. Due to the high source redshift $z_s=2.93$, we have found that the sources of several asymmetric image pairs are too luminous to be just one or a few supergiants even if each star is as bright as $L_{\rm bol}\simeq 10^6\,L_\odot$, but are more likely to be clusters of more than $\sim 10$ bright stars. Our modeling suggests that having $N_\star \sim 10$--$30$ comparably luminous stars shortens the timescale of asymmetry variability to just $\sim 1$ or $2$ yrs. In this case, the asymmetry should even flip its sign several times over six years. This appears, in the case of the brightest image pair, at odds with the consistent sign and degree of asymmetry between optical and IR exposures taken six years apart. For a cluster of $N_\star \gtrsim 100$ comparable stars, microlensing induced asymmetry should be severely diluted, while variability should occur even more rapidly. Since HST images in 2011 and in 2017 were not taken in the same filters, we have not been able to derive a tight constraint on variability, and in fact, variability at some level, however small, is unavoidable. An additional HST epoch in both UV and IR filters would tighten the constraint on any asymmetry component of microlensing origin, giving helpful guidance to upcoming JWST observations.

A population of $\sim 10^6$--$10^8\,M_\odot$ intervening DM subhalos perturbing the magnification symmetry in the proximity of the critical curve is the other viable hypothesis we have considered, for which persistent asymmetries are natural outcomes. Using a reasonable model for subhalo abundance and mass function derived for the CDM paradigm, we have found it plausible that image pairs having more than a $\sim 10\%$ fractional difference in magnifications are common within $\lesssim 0.2\arcsec$ of the cluster critical curve in the case of S1226. 

Precise characterization of the cluster halo of SDSS J1226+2152 and improved substructure modeling will enable us to draw a more robust conclusion on whether asymmetric image pairs neighboring the critical curve in S1226 are caused by substructure lensing. Deeper images taken at many epochs with HST and JWST will help reduce the uncertainty in the flux asymmetry measurement, separate any time-varying component from the persistent component, and rule out possible interloper satellite galaxies or GCs. In particular, in IR filters JWST will have good spatial resolution on par with HST in optical filters.

Nebular lines from HII regions surrounding hot stars, if spatially compact enough, might provide additional evidence for asymmetries insensitive to microlensing thanks to large source sizes. More generally, we expect an increased number of caustic straddling arcs with clumpy star-forming features in the proximity of lensing critical curves to be discovered in the future. Imaging and spectroscopy follow-ups of the source-lens systems will allow us to examine many such magnified image pairs and establish whether asymmetric magnifications are ubiquitous. 

Detecting the population of small-scale subhalos through lensing near caustics will be unprecedented and will complement other efforts to uncover the invisible DM structure on sub-galactic scales, including flux ratio measurements of multiply imaged quasars~\citep{nierenberg2014detection, nierenberg2017probing, 2019MNRAS.487.5721G} and dynamic modeling of Milky Way stellar streams~\citep{johnston2002lumpy, ibata2002substructure, carlberg2009star, 2019arXiv191102663B, 2019ApJ...880...38B}. Thereby, improved constraints will be derived for alternative DM models such as Warm Dark Matter~\citep{kusenko2009sterile, shoemaker2009gravitino, abazajian2017sterile} or ``fuzzy'' DM~\citep{press1990single, PhysRevD.50.3650, goodman2000repulsive, Peebles:2000yy, hu2000fuzzy, 2006PhLB..642..192A, 2014NatPh..10..496S, hui2017ultralight}.

%%%%%%%%%%%%%%%%%%%%%%%%%%%%%%%%%%%%%%%%%%%%%%%%%%%%%%%%%%%%%%%%%%%%
\section*{Acknowledgements}

The authors are grateful to the anonymous referee for important comments that led to substantial improvement of this work. The authors are also thankful for insightful discussions with Eliot Quataert and Kailash Sahu. This work is based on observations made with the NASA/ESA Hubble Space Telescope, obtained from the data archive at the Space Telescope Science Institute. STScI is operated by the Association of Universities for Research in Astronomy, Inc. under NASA contract NAS 5-26555. LD and TV acknowledge the support of John Bahcall Fellowships at the Institute for Advanced Study. AK is supported by the IBM Einstein fellowship. JM has been supported by Spanish Fellowship PRX18/00444, and by the Corning Glass Works Foundation Fellowship Fund. MB acknowledges support by NASA through grant number HST-GO-15378.006-A from the Space Telescope Science Institute, which is operated by AURA, Inc., under NASA contract NAS 5-26555. We acknowledge the use of the publicly available Python package \texttt{Colossus} to carry out cosmological calculations~\citep{diemer2018colossus}.

%%%%%%%%%%%%%%%%%%%%%%%%%%%%%%%%%%%%%%%%%%%%%%%%%%%%%%%%%%%%%%%%%%%%

\appendix

%%%%%%%%%%%%%%%%%%%%%%%%%%%%%%%%%%%%%%%%%%%%%%%%%%%%%%%%%%%%%%%%%%%%
\section{Non-rigid registration}
\label{app:nonrig}
%%%%%%%%%%%%%%%%%%%%%%%%%%%%%%%%%%%%%%%%%%%%%%%%%%%%%%%%%%%%%%%%%%%%

The method of non-rigid registration allows to smoothly deform one surface brightness pattern to match another. In \cite{2019ApJ...880...58K}, this method was applied to the arc in MACS J0416.1-2403. In that case, asymmetries were marginally observed in several slits using deeper co-added HST exposures. Thus, in that case non-rigid registration relied on matching arc surface brightness features seen on both sides of the critical curve to account for imperfections of the ideal fold lens model. The main finding from this method in \cite{2019ApJ...880...58K} was that asymmetries appear more prominent as one examines the portion of the image plane closer to the critical curve, a trend broadly expected from both intracluster microlensing and DM substructure lensing. For S1226, asymmetries are significant in many of the slits we define in \refsec{slits}. Thus, the method of non-rigid registration is naively less reliable. However, we apply it here to see what we can recover. 

\begin{figure}
    \centering
    \includegraphics[scale=0.8]{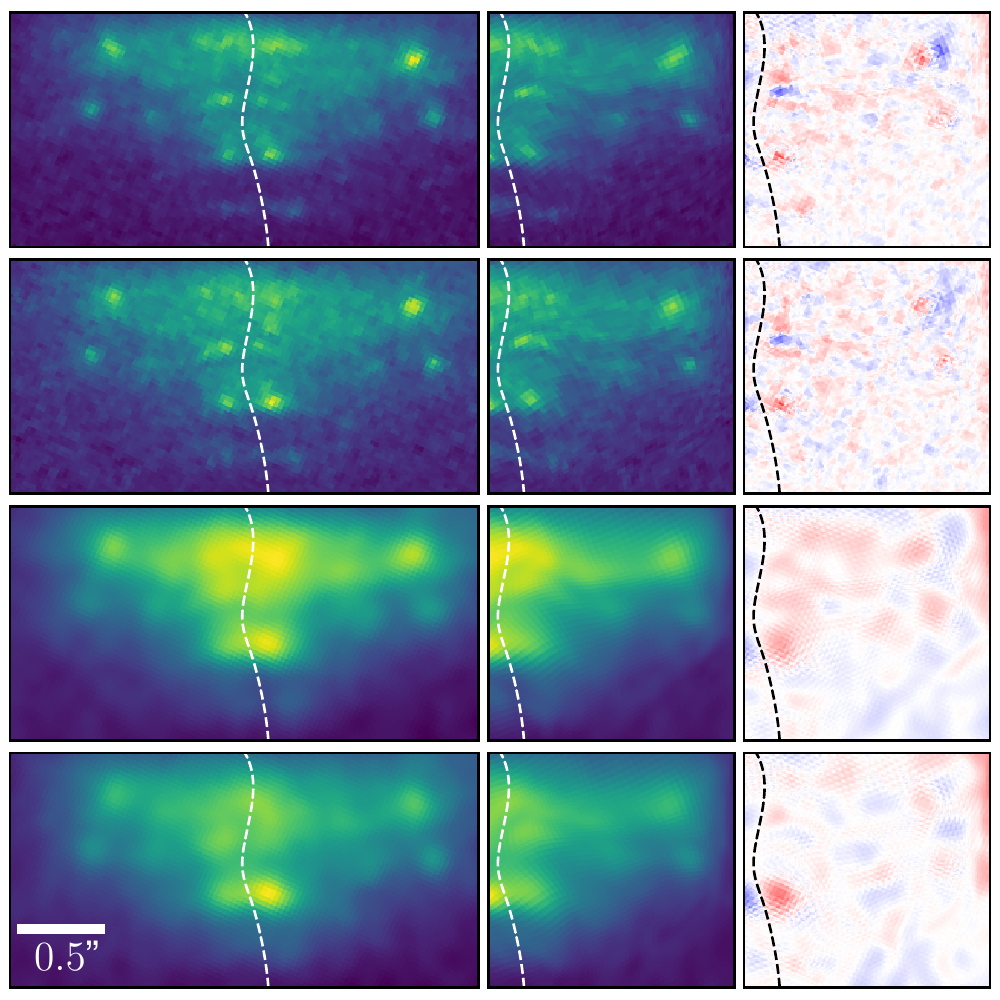}
    \caption{The left column presents the same HST images as in \reffig{slits} for F606W, F814W, F110W and F160W filters. The middle column shows the left part of the arc flipped under non-rigid transformation. The right column shows the surface brightness difference between the left and the middle columns. The white dashed curve indicates the estimated position of the critical curve. }
    \label{fig:nonrig}
\end{figure}

\begin{figure}
    \centering
    \includegraphics[scale=0.83]{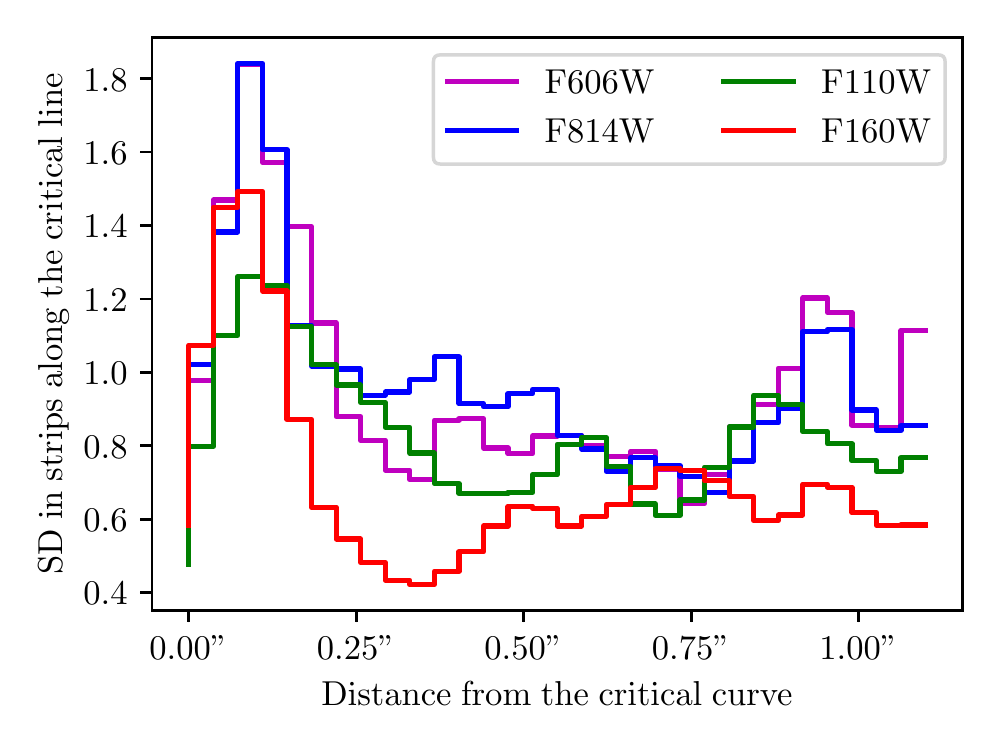}
    \caption{Standard deviation of residuals in the images of
the arc as a function of the angular distance to the critical curve. The standard deviations are calculated in strips parallel to the critical curve using residuals presented in the rightmost column of \reffig{nonrig}. The rise in the standard deviation within $\sim 0.2\arcsec$ may be a sign of intracluster microlensing of individual stars or compact star forming complexes, or substructure lensing from a population of DM subhalos.}
    \label{fig:nonrig_prof}
\end{figure}

The resulting non-rigid transformation is presented in \reffig{nonrig}, where we plot the speculated morphology of the (perhaps perturbed) critical curve as the white dashed curve. The figure highlights many inconsistencies along the critical line, which is in line with what we have found in \refsec{slits}. In \reffig{nonrig_prof}, we show the average amplitude of match residuals as a function of distance to the critical curve, similar to what was done in \cite{2019ApJ...880...58K}. While some surface brightness mismatches are mitigated or eliminated by the non-rigid transformation, others cannot be perfectly cured, especially at smaller distances to the critical curve.

%%%%%%%%%%%%%%%%%%%%%%%%%%%%%%%%%%%%%%%%%%%%%%%%%%%%%%%%%%%%%%%%%%%%
%%%%%%%%%%%%%%%%%%%%%%%%%%%%%%%%%%%%%%%%%%%%%%%%%%%%%%%%%%%%%%%%%%%%

%%%%%%%%%%%%%%%%%%%%%%%%%%%%%%%%%%%%%%%%%%%%%%%%%%%%%%%%%%%%%%%%%%%%
%%%%%%%%%%%%%%%%%%%%%%%%%%%%%%%%%%%%%%%%%%%%%%%%%%%%%%%%%%%%%%%%%%%%

%%%%%%%%%%%%%%%%%%%%%%%%%%%%%%%%%%%%%%%%%%%%%%%%%%%%%%%%%%%%%%%%%%%%
\section{Measuring fluxes of lensed image pairs}
\label{app:analysisdetails}
%%%%%%%%%%%%%%%%%%%%%%%%%%%%%%%%%%%%%%%%%%%%%%%%%%%%%%%%%%%%%%%%%%%%

For our analysis, we use reduced science product images drizzled to a uniform pixel scale of $0.03\arcsec$, one image per filter. The pixel resolution improves upon those of individual exposures, namely a detector pixel scale $0.05\arcsec$ for ACS (F606W and F814W) and $0.128\arcsec$ for WFC3 IR (F110W and F160W).

We use the publicly available software package \texttt{TinyTim}~\citep{krist201120}. To allow sub-pixel centroiding, we need subsampled PSFs output by \texttt{TinyTim}, which neglect the charge diffusion effect~\citep{krist2004position}. Only precisely applicable to individual exposures, those neither include further PSF broadening in drizzled images~\citep{hook2000dithering}. Using stars in the FoV for calibration, we empirically find that convolving \texttt{TinyTim} PSFs with a two-dimensional isotropic gaussian kernel whose sigma equals $0.5$ pixel scale on the detector provides good fits, with $<10\%$ bias in photometry and $<0.01\arcsec$ error in astrometry.

In each filter and for each of the significant asymmetric features in Slits C, D and E, we explore the $\chi^2$ in \refeq{chisq} within a cutout of suitable size which contains the presumptive lensed image pair. We centroid each lensed image within a $0.2\arcsec \times 0.2\arcsec$ search region in (RA, DEC) centering at the pixel of peak flux. Some of the relevant parameters adopted in our analysis can be found in \reftab{chisqdetailtab}.

We adopt simple and reasonable parametrized models for the diffuse background based on the surface brightness properties within each cutout. For the image pairs in Slits C and E, we adopt the \texttt{gradient} model which is simply a two-dimensional linear function with 3 coefficients:
\begin{align}
    \label{eq:gradientbkg}
    S_0 + S_1\,{\rm RA} + S_2\,{\rm DEC}.
\end{align}
This form captures any smooth ICL and sky backgrounds.

The brighter image pair in Slit D happen to be located at a boundary separating two distinct sections of the arc with high and low surface brightness, respectively. The arc is bright enough that ICL and sky backgrounds are unimportant. Near the lensing critical curve, this boundary is physically constrained to be parallel to the direction of arc elongation. We therefore introduce the \texttt{smooth step} model for the diffuse background, which is the sum of a uniform component and a one-dimensional profile describing a smooth transition in the surface brightness:
\begin{align}
    \label{eq:smoothstepbkg}
    S_0 + S_1\, \tanh\left( \frac{\cos\theta\,{\rm RA} - \sin\theta\,{\rm DEC} - \Delta}{b} \right).
\end{align}
The \texttt{smooth step} model has a total of 5 parameters, which include two constants $S_0$ and $S_1$, and 3 other parameters $(\theta, \Delta, b)$ characterizing the orientation, position, and width of the surface brightness transition respectively.

\begin{figure*}
    \centering
    \includegraphics[scale=0.27]{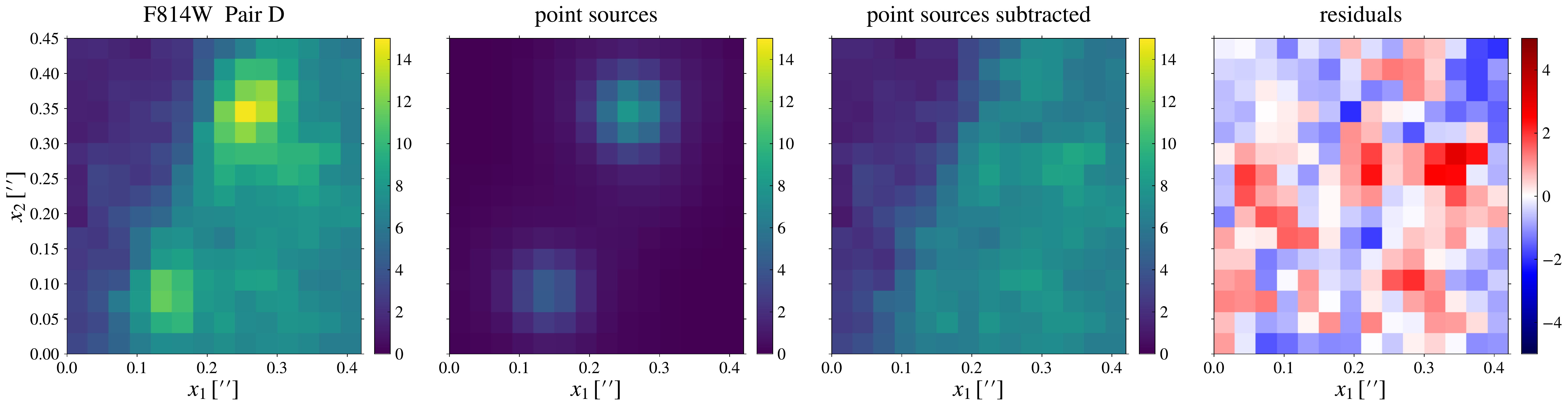}
    \includegraphics[scale=0.27]{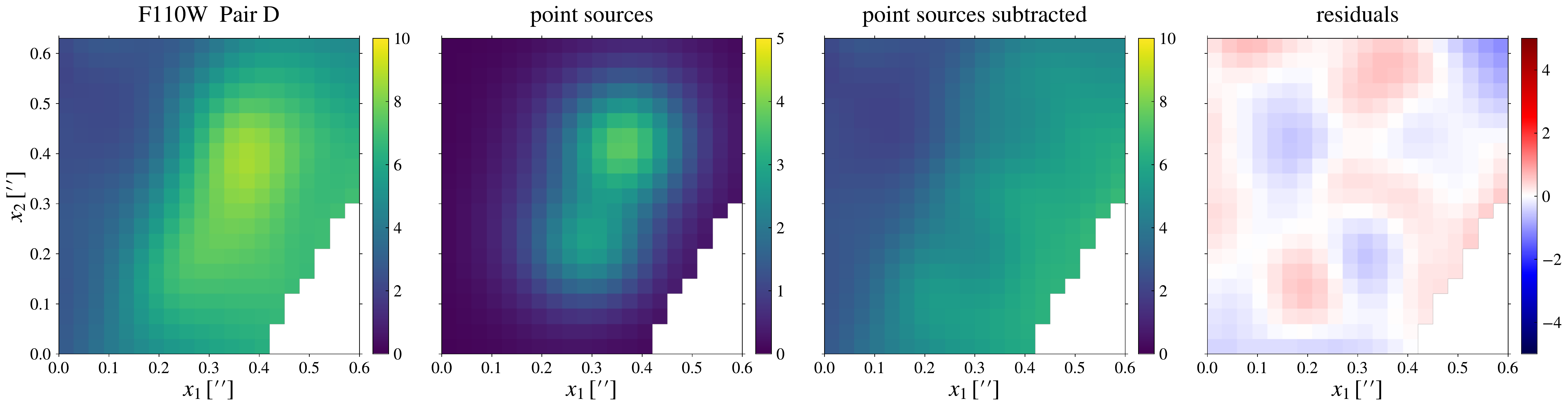}
    \caption{Example fitting results for the image pair in Slit D in the F814W filter (upper row) and in the F110W filter (lower row). The maximum likelihood solutions are shown. The four columns, from left to right, show the drizzled data ($0.03\arcsec$/pixel), the fitted lensed image pair modeled as a pair of PSFs, the data after the image pair are subtracted, and residuals, respectively. For panels in the first three columns, fluxes are color coded in units of ${\rm nJy}$ per pixel. For panels showing the residuals, fluxes are color coded in units of $\sigma_F$, whose values can be found in \reftab{chisqdetailtab}. Note the different cutout sizes used for F814W and F110W.}
    \label{fig:fit_results_example}
\end{figure*}

For the inverse variance weights $\sigma_i$'s in \refeq{chisq}, we adopt the white noise approximation by setting all $\sigma_i$'s to be a constant $\sigma_F$, which models fit residuals due to both flux measurement errors and uncertainties in the intrinsic surface brightness inhomogeneities on the arc. Its value is tuned such that for the maximum likelihood solution the $\chi^2$ per effective degree of freedom is around unity. In the two optical filters F606w and F814W, the total number of effective degrees of freedom is taken to be the total number of cutout pixels $N_{\rm pix}$ minus the total number of fitting parameters $n_{\rm dim}$. The drizzled $0.03\arcsec$/pixel images for F110W and F160W are oversampled, which causes residuals to show correlation across many pixels (see bottom rightmost panel of \reffig{fit_results_example}). We estimate a factor of 4 for oversampling in resolution, and hence reduce the effective number of pixels by a factor of 16 in those two filters.

As an example, in \reffig{fit_results_example} we show results of our aforementioned fitting procedure for the most prominent asymmetric image pair feature in Slit D, in one optical filter F814W and one IR filter F110W. The fitting residuals for F814W show only moderate departure from white noise, mainly due to the presence of secondary underlying compact sources on the arc. We do not expect the white noise approximation leads to significant mis-estimation of the errorbars. The fitting residuals for F110W clearly exhibit correlation as expected from oversampling, justifying the reduction in the effective number of pixels used to tune $\sigma_F$.

For a cross check, we re-perform the analysis using HST images of individual visits (two visits each for F606W and F814W with $0.05\arcsec$/pixel; one visit each for F110W and F160W with $0.128\arcsec$/pixel) produced by the standard reduction pipeline and downloadable from the Mikulsky Archive for Space Telescopes (\url{https://mast.stsci.edu/portal/Mashup/Clients/Mast/Portal.html}), properly treating astrometric misalignment between images taken from different visits. The results obtained are in good agreement with what we conclude with our own drizzled images.

%%%%%%%%%%%%%%%%%%%%%%%%%%%%%%%%%%%%%%%%%%%%%%%%%%%%%%%%%%%%%%%%%%%
%%%%%%%%%%%%%%%%%%%%%%%%%%%%%%%%%%%%%%%%%%%%%%%%%%%%%%%%%%%%%%%%%%%

%%%%%%%%%%%%%%%%%%%%%%%%%%%%%%%%%%%%%%%%%%%%%%%%%%%%%%%%%%%%%%%%%%%%
\begin{table*}
	\centering
	\caption{Detailed choices for $\chi^2$ minimization of \refeq{chisq} for different asymmetric features in various filters. $N_{\rm pix}$ is counted without applying over-sampling reduction. $n_{\rm dim}$ is the total number of dimensions of the parameter space to explore.}
	\label{tab:chisqdetailtab}
	\begin{tabular}{ccccccc}
		\hline
		\hline
		image pair & HST filter & $N_{\rm pix}$ & $n_{\rm bkg}$ & diffuse background model & $n_{\rm dim}$ & $\sigma_F\,[{\rm nJy}]$ \\
		\hline
		\hline
		Slit C & & & & & & \\
		\hline
		& F606W & 195 & 2 & \texttt{gradient} & 9 & 0.60 \\
		& F814W & 188 & 2 & \texttt{gradient} & 9 & 0.80 \\
		& F110W & 264 & 2 & \texttt{gradient} & 9 & 1.20 \\
		& F160W & 264 & 2 & \texttt{gradient} & 9 & 1.70 \\
		\hline
		\hline
		Slit D & & & & & & \\
		\hline
		& F606W & 210 & 2 & \texttt{smooth step} & 11 & 0.50 \\
		& F814W & 210 & 2 & \texttt{smooth step} & 11 & 0.65 \\
		& F110W & 383 & 2 & \texttt{smooth step} & 11 & 1.00 \\
		& F160W & 383 & 2 & \texttt{smooth step} & 11 & 1.60 \\
		\hline
		\hline
		Slit E & & & & & & \\
		\hline
		& F606W & 339 & 2 & \texttt{gradient} & 9 & 0.40 \\
		& F814W & 253 & 2 & \texttt{gradient} & 9 & 0.54 \\
		& F110W & 406 & 2 & \texttt{gradient} & 9 & 0.60 \\
		& F160W & 406 & 2 & \texttt{gradient} & 9 & 1.50 \\
		\hline
	\end{tabular}
\end{table*}
%%%%%%%%%%%%%%%%%%%%%%%%%%%%%%%%%%%%%%%%%%%%%%%%%%%%%%%%%%%%%%%%%%%%

%%%%%%%%%%%%%%%%%%%%%%%%%%%%%%%%%%%%%%%%%%%%%%%%%%%%%%%%%%%%%%%%%%%%
\section{Magnification asymmetry across critical curve}
\label{app:magasym}
%%%%%%%%%%%%%%%%%%%%%%%%%%%%%%%%%%%%%%%%%%%%%%%%%%%%%%%%%%%%%%%%%%%%

In this Appendix, we estimate the degree of uneven fluxes between a close pair of highly magnified images on both sides of the critical curve. We aim to derive the parametric dependence on the perturber parameters. 

Consider a pair of images at image plane positions $x_{1,i}$ and $x_{2, i}$. Demanding that they map to the same source point, we obtain from the lens equation
\begin{align}
    x_{2, i} - x_{1, i} = \alpha_i(x_2) - \alpha_i(x_1),
\end{align}
where $\alpha_i(x)$ is the deflection field as a function of the two-dimensional image plane vector $x$. Let us write $x_{1,i} = x_{0, i} - \Delta x_i$ and $x_{2,i} = x_{0, i} + \Delta x_i$, where $x_{0, i}$ is the midpoint between the image pair. 

Assuming that $\alpha_i(x_j)$ is smooth, we Taylor expand in powers of $\Delta x_i$:
\begin{align}
    \Delta x_i & = \left.\left[ \nabla_j\,\alpha_i \right]\right|_{x_0}\,\Delta x_j + \frac{1}{6}\,\left.\left[ \nabla_l\,\nabla_k\,\nabla_j\,\alpha_i \right]\right|_{x_0}\,\Delta x_l\,\Delta x_k\,\Delta x_j \nonumber\\
    & + \mathcal{O}\left[\Delta x^{5}\right],
\end{align}
where $\nabla_i$ denotes image-plane derivative with respect to $x_i$, and $\left.[\cdots]\right|_{x_0}$ stands for evaluation at $x_0$. Ignoring terms at cubic and higher orders, the Jacobian matrix $J_{ij}(x)=\delta_{ij} - \nabla_i\,\alpha_j(x)$ at $x_0$ satisfies $\left.[J_{ij}]\right|_{x_0}\,\Delta x_j \approx 0$. This means that $\Delta x_i$ is an eigenvector with a vanishing eigenvalue.

The inverse of the signed magnification is
\begin{align}
    1/\mu(x) = {\rm det}\left[ \delta_{ij} -  \nabla_i\,\alpha_j(x)\right],
\end{align}
where we again Taylor expand around $x=x_0$:
\begin{align}
\label{eq:JijTaylorExpand}
    \delta_{ij} -  \nabla_i\,\alpha_j(x_{1,2}) & = \left.[J_{ij}]\right|_{x_0} \pm \frac12\,\left.\left[ \nabla_k\,\nabla_j\,\alpha_i \right]\right|_{x_0}\,\Delta x_k \\
    & - \frac16\,\left.\left[ \nabla_l\,\nabla_k\,\nabla_j\,\alpha_i \right]\right|_{x_0}\,\Delta x_l\,\Delta x_k + \mathcal{O}\left[\Delta x^{3}\right]. \nonumber
\end{align}
In a suitably oriented coordinate system, $\Delta x_i$ aligns with the first axis, and the only nonzero element of $\left.[J_{ij}]\right|_{x_0}$ is the 22 element being order unity.

If we neglect terms of $\mathcal{O}\left[\Delta x^2\right]$ in \refeq{JijTaylorExpand}, the determinant is at the leading order proportional to the $\mathcal{O}\left[\Delta x\right]$ term in the 11 element of the full Jacobian matrix. Since this term has equal magnitudes but opposite signs at both image positions $x_1$ and $x_2$, we recover the familiar result that a pair of images across an ideal fold are equally magnified.

Departure from perfect magnification symmetry is induced by two types of correction in \refeq{JijTaylorExpand}: (i) the $\mathcal{O}\left[\Delta x^2\right]$ term in the 11 element; (ii) the product of $\mathcal{O}\left[\Delta x\right]$ terms from two matrix elements. Having the same signs at $x_1$ and at $x_2$, these correct the leading magnification factor in opposite directions at the two images and hence generate asymmetry.

The lowest order derivative of the deflection field is usually determined by the large-scale lens, while the higher order derivatives can be dominated by small perturber lenses. By definition, $\Delta x$ is on the order of the image separation $\Delta\theta$. Neglecting numerical prefactors, we assume $\left.\left[ \nabla_k\,\nabla_j\,\alpha_i \right]\right|_{x_0}$ is of order $d$ from the large-scale mass distribution. We assume that a perturber lens dominates the derivative at the next order,
\begin{align}
    \left.\left[ \nabla_l\,\nabla_k\,\nabla_j\,\alpha_i \right]\right|_{x_0} \sim \theta^2_p/b^4,
\end{align}
as long as the angular impact parameter $b$ is larger than the characteristic Einstein radius $\theta_p$ of the perturber. We also require $b \gtrsim \Delta \theta$ so that Taylor expansion in $\Delta x$ is justified. Fractional correction to magnification from terms of type (i) is therefore
\begin{align}
    |a_\mu| \sim \left(\theta^2_p\,\Delta\theta\right)/\left(d\,b^4\right).
\end{align}
The fractional correction to magnification from terms of type (ii) is of order $d\,\Delta\theta$, independent of the perturber. This is comparable to the inverse of the leading order magnification factor. For image pairs of interest in this work, the magnification factor is as large as $\gtrsim 100$. Hence, these terms only induce insignificant fractional magnification asymmetry at (sub-)percent levels.  

%%%%%%%%%%%%%%%%%%%% REFERENCES %%%%%%%%%%%%%%%%%%
\bibliographystyle{mnras}
\bibliography{refs.bib,refs2.bib,refs3.bib}

% Don't change these lines
% \bsp	% typesetting comment
\label{lastpage}
\end{document}